\documentclass{aa}  
\usepackage[varg]{txfonts}
\usepackage{graphicx}
\usepackage{array}
\usepackage{lscape}
\usepackage{txfonts}
\usepackage[normalem]{ulem}
\usepackage{pdflscape}
\usepackage{color}
\usepackage{rotating}
\usepackage{supertabular}
\usepackage{mathrsfs}
\usepackage{stfloats}
\usepackage[flushleft]{threeparttable}
\usepackage{multirow}
\usepackage{lscape}
\usepackage[table]{xcolor}
\usepackage{longtable}

\fancypagestyle{mylandscape}{
\fancyfoot{
\makebox[\textwidth][r]{
  \rlap{\hspace{1.5cm}
    \smash{
      \raisebox{13.6cm}{
        \rotatebox{90}{\thepage}}}}}}
        }
\usepackage{amsmath}
\let\oldAA\AA
\renewcommand{\AA}{\text{\normalfont\oldAA}}

\usepackage{hyperref}
\begin{document}

\title{Astro+ database}
\subtitle{I. Description and first results}

\author{ K. R\"ubke$^{1,2}$,
 A. Marco$^{1,3}$,
 I. Negueruela$^{2,3}$,
 A. Herrero$^{4,5}$,
 S. Sim\'{o}n-D\'{\i}az$^{4,5}$,
 H. M. Tabernero$^{7,8}$
 \&  L. R. Patrick$^{6}$}

\institute{$^{1}$Departamento de FÃ­sica, IngenierÃ­a de Sistemas y TeorÃ­a de la SeÃ±al, Universidad de Alicante, Carretera de San Vicente s/n, E03690, San Vicente del Raspeig, Spain e-mail: klaus.rubke@ua.es
\\
$^{2}$Departamento de FÃ­sica, Facultad de Ciencias, Universidad de Alicante
\\
${^3}$Instituto Universitario de InvestigaciÃ³n InformÃ¡tica. Universidad de Alicante\\
${^4}$Instituto de AstrofÃ­sica de Canarias, VÃ­a LÃ¡ctea s/n, E38200, La Laguna, Tenerife, Spain
\\
${^5}$Universidad de La Laguna, Departamento de AstrofÃ­sica, E38206 La Laguna, Tenerife, Spain
\\
${^6}$Centro de AstrobiologÃ­a (CAB), CSIC-INTA, Carretera de Ajalvir km4, 28850 TorrejÃ³n de Ardoz, Madrid, Spain
\\
${^7}$Institut d'Estudis Espacials de Catalunya (IEEC), Edifici RDIT, Campus UPC, 08860 Castelldefels (Barcelona), Spain
\\
${^8}$Institut de CiÃ¨ncies de l'Espai (ICE, CSIC), Campus UAB, c/ de Can Magrans s/n, 08193 Cerdanyola del VallÃ¨s, Barcelona, Spain
\\
}

   \date{Received  ; accepted  }

\abstract
   {} 
   {The vast amounts of spectroscopic data for massive stars provided by previous and existing instruments on ground-based and space-based telescopes have saturated our capability to process them by human inspection routines. Consequently, there is a pressing need for fully automatic machine-assisted tools to help handle incoming data. To this end, we present the development of a massive star spectroscopic interactive database, Astro+. We aim to provide users with a reliable, versatile, and user-friendly platform that will be significant for understanding massive stars and set an important precedent for future open-access astrophysical research.}
   {This tool allows authorized users to upload their own spectra and by using the fully automated tool HiLineThere, a Python-based program, it can homogeneously derive basic stellar parameters, such as $v_{\mathrm{rad}}$, $v \sin\, i$, $T_{\mathrm{eff}}$, and $\log \, g$ for massive OB-type stars, using a solar-metallicity grid of FASTWIND models, and $v_{\mathrm{rad}}$, $[\mathrm{Fe/H}]$, $T_{\mathrm{eff}}$, and $\log \, g$ for red supergiant stars, in a completely autonomous way.}
   {Here we present the first results of the tool HiLineThere on optical spectra for OB-type and red supergiants. We compare the output of our analysis for OB-type stars with literature values for a large sample of well-studied objects, finding differences within the expected error ranges: $T_{\mathrm{eff}} \sim 800$\,K, $\log g \sim 0.1$\,dex, and $v \sin i \sim 10$\,km\,s$^{-1}$. Preliminary tests on early-B stars also show consistent results during the transition to lower temperatures. For red supergiants, we find differences within $7$\,km\,s$^{-1}$ in $v_{\mathrm{rad}}$ and $\sim 300$\,K in $T_{\mathrm{eff}}$ for two test samples.}
   {}
    \titlerunning{Astro+}
\authorrunning{R\"ubke et al.}
\maketitle
\nolinenumbers

\section{Introduction}

Massive stars play a fundamental role in the evolution and formation of their host galaxies. Although less than 1 percent of stars born are estimated to be massive \citep{Kroupa2002}, their impact far exceeds their scarcity. In their youth, they are highly energetic, with fully radiative atmospheres that expel large amounts of mass into the interstellar medium (ISM) through stellar winds. This process perturbs their environment and can even compress surrounding gas into overdense regions, triggering new star formation. In their final evolutionary stage, most commonly as red supergiants (RSGs), these stars end their lives in supernova explosions, releasing their outer layers into the ISM, enriching it with heavy elements, increasing galactic metallicity, and leaving behind either a neutron star or a stellar black hole, depending on their initial mass.

Identifying and studying massive stars has been challenging due to their scarcity, as their lifespans are much shorter than those of lower-mass stars. As a consequence, in the Milky Way, they are preferentially located close to their birthplaces, in regions of dense gas and dust that obscure them from view. Many become detectable only after they have blown out the ionized gas or dynamic interactions have expelled them from their natal environment \citep{Zinnecker2007}. Even then, photometric identification of massive stars is complicated by extinction, either from the molecular cloud where they were born or from the ISM, and varying distances, which can misplace them in color-color or color-magnitude diagrams \citep{Massey1995,Negueruela2023}.

Spectroscopy is a powerful tool for deriving stellar parameters by analyzing line ratios and intensities \citep[e.g.,][]{maiz2026}, or entire spectral ranges compared to theoretical models \citep{Puls2026,Gray2022,Martins2018,ssimon2020b}. This approach provides key parameters such as temperature, gravity, metallicity, rotational velocity, and radial velocity of the star. However, modeling stellar atmospheres is complex and requires detailed physical considerations. The most advanced model families focus on specific temperature ranges, as the physics involved may be different or require different treatment. In particular, the atmospheres of young massive stars are characterized by nonlocal thermodynamic equilibrium (NLTE) conditions, and complex physics must be considered for accurate spectral synthesis. Several codes have been developed to model stellar atmospheres, each with their own strengths and weaknesses. CMFGEN \citep{hillier2001}, PoWR \citep{Grafener2002}, WM-basic \citep{Pauldrach1986}, and FASTWIND \citep{santolaya97} are NLTE codes that include the effects of stellar winds, whereas TLUSTY \citep{tlusty2003} and ATLAS9 \citep{CastelliKurucz2003} are static, plane-parallel codes, the latter assuming local thermodynamic equilibrium (LTE). A recent comparison of atmosphere codes within the X-Shooting ULLYSES collaboration found that CMFGEN, PoWR, and FASTWIND yield broadly consistent results for O-type stars \citep{Sander2024}. Among the wind codes, FASTWIND achieves a very good balance between accuracy and computational efficiency in the range of hot massive stars. Likewise, codes such as PHOENIX \citep{Husser2013} and MARCS \citep{Gustafsson2008} accurately model lower-temperature stars, including high-mass objects such as cool supergiant stars.

Over the years, numerous catalogs have compiled extensive spectral datasets of massive stars, which are now publicly available. Among them, we can cite OWN \citep{Barba2010,Barba2017,Barba2026}, VFTS \citep{Evans2011}, GES \citep{Blomme2022}, GOSSS \citep{Maiz2011}, and IACOB \citep{ssimon2011,ssimon2015,ssimon2020}. To take advantage of these resources, we have developed a centralized and standardized database to store, organize, and facilitate the analysis of massive star spectra for the astronomical community. Due to the wide range of data collected from various telescopes and instruments, the database is designed to support multiple spectral data format inputs, providing a unified and adaptable repository for heterogeneous spectroscopic datasets. In addition, during the upload of any spectrum to the database, astronomical application programming interfaces (APIs), such as Aladin, Simbad, and the Gaia archive, are used to systematically obtain all relevant catalog data, supporting the identification, validation, and spectroscopic analysis of the massive star spectrum.

In this work, we introduce Astro+\footnote{\url{https://astroplus.ua.es/}}, the first multiwavelength database dedicated to the spectra of massive stars, which can integrate data from X-ray to far-infrared wavelengths, although in this current version it is focused on the analysis of optical spectra only.  At present, the application contains over 4\,000 stellar spectra, mostly OB and RSG spectra drawn from several public catalogs (i.e., OWN, IACOB, and GOSSS), together with data from our own observing programs, offering the scientific community a user-friendly interface to contribute additional data in ASCII or FITS formats. Beyond storage, Astro+ is designed to perform a fully automated analysis of these optical spectra. Traditionally, identifying and characterizing any stellar spectrum requires the expertise of an astrophysicist. An expert in spectral analysis of massive stars should perform various tasks, such as spectrum normalization, the selection of diagnostic lines, artifact removal, correction for radial velocity, and the determination of rotational broadening and macroturbulence. Examples of this approach can be found in numerous works \citep[e.g.,][]{Martins2005, Repolust2005, Sabin2014, RamirezA2017, Dorda2018, holgado2018, tabernero2018}. However, with the number of OB star spectra now exceeding several thousand, plus thousands more for RSGs, we have already reached the limit of what can be handled with traditional methods.

Moreover, the immediate arrival of large-scale spectroscopic surveys that make use of multiobject spectrographs, such as WEAVE \citep{Jin2024}, will soon add tens of thousands of spectra in both the OB and RSG domains. Astro+ addresses this challenge by providing uniform processing of the spectra and a homogeneous, fully automated spectral analysis. Using a variety of programming tools and routines, we present HiLineThere, an autonomous pipeline capable of performing classification and stellar parameter determination for massive stars. By relying on a hierarchical decision logic, the code automatically routes spectra through specialized modeling paths to cover the full OBAFGKM range: FASTWIND models for hot OB stars (the blue path), an intermediate transition using KURUCZ/ATLAS9 models (the yellow path), and an implementation of \textsc{SteParSyn} \citep{tabernero2022} utilizing MARCS models for cool, late-type stars (the red path).

In this paper, we present the functionality of the Astro+ database, including data upload and the automated spectral analysis framework. We also validated our results against literature benchmarks for a representative sample spanning from young, massive O-type stars and early B-type stars, down to evolved cool RSGs. The samples presented and analyzed in this first paper are Galactic; the extension of the analysis to other metallicity regimes is planned for future versions (Sect.~\ref{sec:conclusions}). The paper is organized as follows. Section~\ref{Science} details the operation of the HiLineThere tool and its classification logic; Section~\ref{ResultDiscussion} presents comparisons with the existing literature, and Section~\ref{sec:conclusions} summarizes our conclusions.

\section{Automatic stellar classification and parameter determination with HiLineThere}\label{Science}
Similarly to standard web applications, Astro+ is made up of multiple components, each with its own background processes. The Web interface manages the ingestion, processing, and visualization of spectral data, providing users with intuitive tools to upload and review their stellar spectra. For further details and functionalities, see the Appendix \ref{sec:database}.

The ability to compile stellar spectra into a single repository offers an excellent starting point for developing homogeneous analysis tools. Although fully automating spectrum identification, characterization, and stellar parameter determination may seem ambitious, this goal has been achieved with the development of HiLineThere, a fully automated program that applies artificial intelligence techniques to tasks traditionally requiring human expertise. The tool classifies stellar spectra and determines the corresponding atmospheric parameters by integrating state-of-the-art diagnostic methods. It follows a structured workflow, as illustrated in Fig.~\ref{fig:WF_HLT}.

\begin{table}[ht]
\scriptsize
\caption{Diagnostic lines used by Astro+ for early-type stars.   \label{table:lines}}
\begin{center}
\begin{tabular}[t]{lllc}\hline
Line      & Lambda  & $V_{\mathrm r}$  &$V \sin\, i$ / $\zeta$\\\hline\hline
\rowcolor{blue!15}
\ion{H}{}$\alpha$   & 6562.82 &   & \\\rowcolor{blue!15}
\ion{H}{}$\beta$    & 4861.33 & X  & \\\rowcolor{blue!15}
\ion{H}{}$\gamma$   & 4340.47 & X  & \\\rowcolor{blue!15}
\ion{H}{}$\delta$   & 4101.74 &  X & \\\rowcolor{blue!15}
\ion{H}{}$\epsilon$ & 3970.08 &  X & \\\rowcolor{blue!15}
\ion{He}{i} 4026   & 4026.19 & X & 15\\\rowcolor{blue!15}
\ion{He}{i} 4121   & 4120.82 &  & \\\rowcolor{blue!15}
\ion{He}{i} 4143   & 4143.76 &  & \\\rowcolor{blue!15}
\ion{He}{i} 4387   & 4387.93 & X & \\\rowcolor{blue!15}
\ion{He}{i} 4471   & 4471.50 & X &  9\\\rowcolor{blue!15}
\ion{He}{i} 4713   & 4713.21 & X  &7 \\\rowcolor{blue!15}
\ion{He}{i} 4922   & 4921.93 &   &14 \\\rowcolor{blue!15}
\ion{He}{i} 5015   & 5015.67 & X  & 8 \\\rowcolor{blue!15}
\ion{He}{i} 5875   & 5875.66 & X &10 \\\rowcolor{blue!15}
\ion{He}{i} 6678   & 6678.15 &   & \\\rowcolor{blue!15}
\ion{He}{ii} 4200  & 4199.90 &   & 11\\\rowcolor{blue!15}
\ion{He}{ii} 4541  & 4541.67 & X & 13\\\rowcolor{blue!15}
\ion{He}{ii} 4686  & 4685.85 &   & \\\rowcolor{blue!15}
\ion{He}{ii} 5411  & 5411.61 & X &  12\\\rowcolor{gray!15}
\ion{O}{ii} 4661   & 4661.63 &   & \\\rowcolor{gray!15}
\ion{O}{ii} 4710   & 4710.01 &   & \\\rowcolor{gray!15}
\ion{O}{iii} 5592  & 5592.37 & X & 1\\\rowcolor{gray!15}
\ion{Si}{ii} 4128  & 4128.07 &   & \\\rowcolor{gray!15}
\ion{Si}{ii} 4131  & 4130.89 &   & \\\rowcolor{gray!15}
\ion{Si}{ii} 6347  & 6347.11 &   & \\\rowcolor{gray!15}
\ion{Si}{ii} 6371  & 6371.37 &  X & \\\rowcolor{gray!15}
\ion{Si}{iii} 4552 & 4552.62 & X & 2\\\rowcolor{gray!15}
\ion{Si}{iv} 4089  & 4088.85 &   & 5 \\\rowcolor{gray!15}
\ion{Si}{iv} 4116  & 4116.10 & X & 6 \\\rowcolor{gray!15}
\ion{Si}{iv} 4629  & 4628.62 &   & \\\rowcolor{gray!15}
\ion{Si}{iv} 4631  & 4631.24 &   & \\\rowcolor{gray!15}
\ion{Si}{iv} 4638  & 4638.28 &   & \\\rowcolor{gray!15}
\ion{Si}{iv} 6668  & 6667.56 &   & \\\rowcolor{gray!15}
\ion{Si}{iv} 6701  & 6701.21 &   & \\\rowcolor{gray!15}
\ion{Mg}{ii} 4481  & 4481.15 & X & 4\\\rowcolor{gray!15}
\ion{C}{ii} 4267   & 4267.24 & X & 3\\\rowcolor{gray!15}
\ion{N}{ii} 4433   & 4432.74 &  & \\ \rowcolor{gray!15}
\ion{N}{ii} 4447   & 4447.03 &  & \\ \rowcolor{gray!15}
\ion{N}{ii} 6548   & 6548.05 &  & \\ \rowcolor{gray!15}
\ion{N}{ii} 6583   & 6583.45 &  & \\ \rowcolor{gray!15}
\ion{N}{iii} 4379  & 4379.11 & X & \\ \rowcolor{gray!15}
\ion{N}{iii} 4511  & 4510.91 & X & \\ \rowcolor{gray!15}
\ion{N}{iii} 4515  & 4514.86 & X & \\ \rowcolor{gray!15}
\ion{N}{iii} 4534  & 4534.58 &   & \\ \rowcolor{gray!15}
\ion{N}{iv} 4058   & 4057.76 &  & \\ \rowcolor{gray!15}
\ion{N}{iv} 6381   & 6380.77 &  & \\ \rowcolor{gray!15}
\ion{N}{v} 4603    & 4603.73 &  & \\ \rowcolor{gray!15}
\ion{N}{v} 4620    & 4619.98 &  & \\ \rowcolor{gray!15}
\ion{Fe}{ii} 4233  & 4233.16 &  & \\ \rowcolor{gray!15}
\ion{Fe}{ii} 4508  & 4508.28 &  & \\ \rowcolor{gray!15}
\ion{Fe}{ii} 4583  & 4583.84 &  & \\ \rowcolor{gray!15}
\ion{Fe}{ii} 4716  & 4716.20 &  & \\ \rowcolor{gray!15}
\ion{Fe}{ii} 4923  & 4923.92 &  & \\ \rowcolor{gray!15}
\ion{Fe}{ii} 5018  & 5018.44 &  & \\ 
\ion{Ca}{ii} 3934  & 3933.66 &  & \\ 
\ion{Na}{i}  5889  & 5889.95 &  & \\
\ion{Na}{i}  5895  & 5895.92 &  & \\\hline \hline
\end{tabular}
\end{center}
\tablefoot{
Diagnostic lines serve different purposes based on their characteristics. Lines marked with an "X" are used to determine $V_{\mathrm r}$, while those used to calculate $V \sin\,i$ and $\zeta$ are numbered according to their priority. Lines with a light blue background are specifically employed to determine parameters for early-type OB stars, and gray lines, although not used for final parameters like temperature, are utilized for determining $V \sin\, i$, macroturbulence, and radial velocity. Finally, lines without any background color are primarily used for additional diagnostics, such as cleaning contaminations or addressing blending issues.}
\end{table}

 HiLineThere works under the sole assumption that the spectra fed to it correspond to massive stars. Beyond this, the tool is fully capable of autonomously classifying and characterizing spectra of both young and evolved massive stars, with no other input that the spectra themselves. Initially, the routine tries to identify typical lines present in OB-type stars. The process not only searches for the lines, but, if it finds them, determines basic parameters such as the radial velocity ($V_{\mathrm r}$), the projected rotational velocity $V \sin\,i$ and the macroturbulence $\zeta$. At present, the workflow treats every input spectrum as that of a single star. Single-lined spectroscopic binaries (SB1) are therefore analyzed as single stars with a shifted radial velocity, which does not affect the determination of the atmospheric parameters. Double-lined systems with a faint secondary are likewise fitted as apparently single stars, whereas a double-lined binary with components of comparable brightness would be fitted with a single set of parameters, showing up as a poor or degenerate fit rather than being deblended. A dedicated treatment of such SB2 systems is left for future work.

At this stage, the decision-making process follows a hierarchical logic: first, the routine searches for the presence of the Balmer series. If these lines are absent or poorly determined, the star is immediately classified as a cool candidate and follows the red path. In this branch, a sub-program called HiBandThere determines $V_{\mathrm r}$ by cross-correlating the spectrum with 64 cool stellar models, selecting the model which minimizes the residuals. The best-matching velocity is then used in the final analysis step, where \textsc{SteParSyn} \citep{tabernero2022}, a Bayesian code designed to infer the atmospheric parameters of late-type stars, derives the atmospheric parameters.

If, however, the Balmer series is well-defined, a second assessment (a ``minifilter'') estimates if the temperature is above or below a threshold of approximately 15\,000~K. This minifilter performs a preliminary check using a representative subset of approximately 1\,500 FASTWIND models. This sub-grid is strategically selected to cover the core $T_{\mathrm{eff}}$ and $\log\,g$ parameter space, providing a fast first-order approximation. It is important to note that this stage serves only as a diagnostic gatekeeper for path selection.

\begin{figure}
\centering
\resizebox{\hsize}{!}{\includegraphics{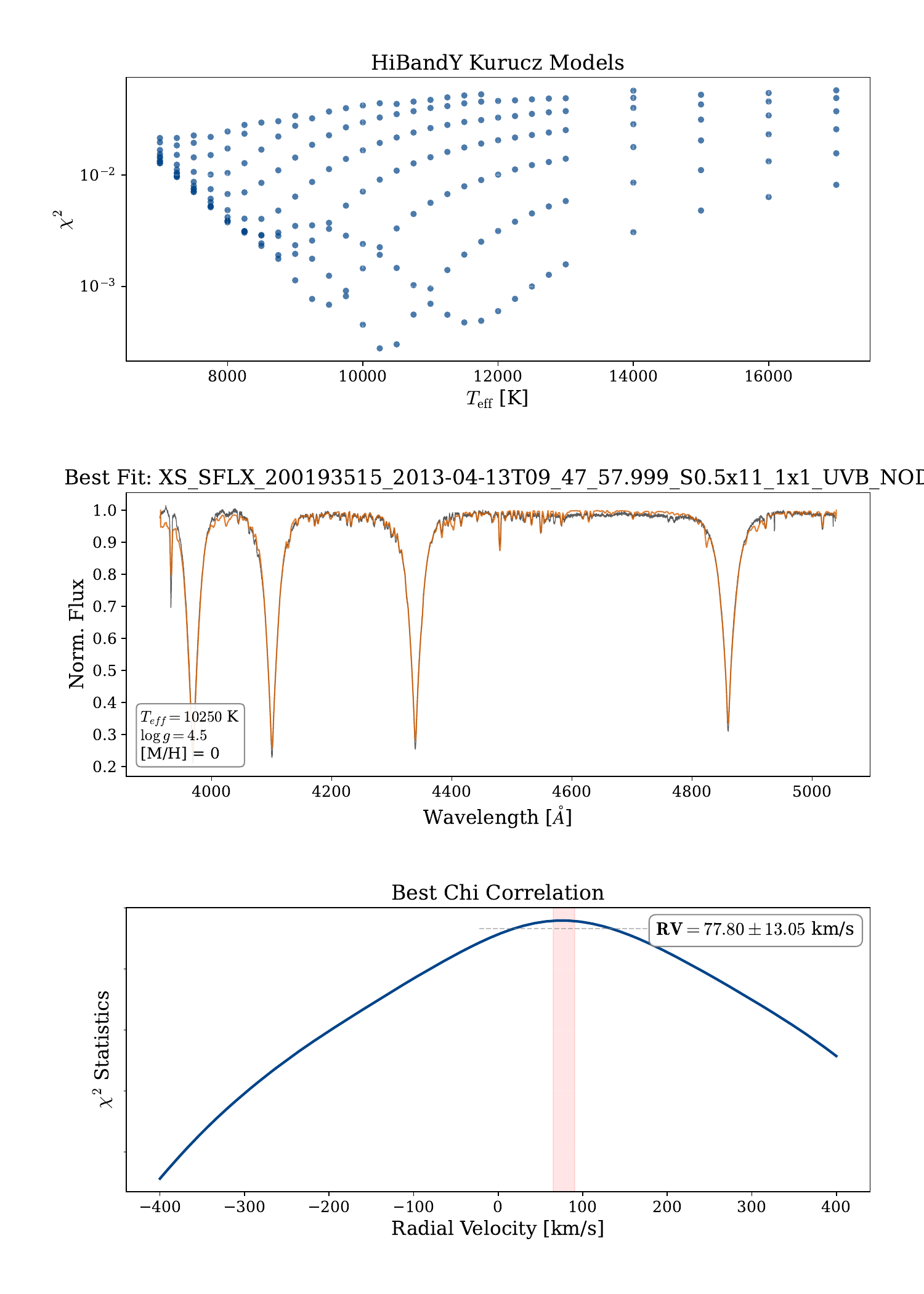}}
\caption{\label{fig:hibandtherey} Example output from the HiBandThereY subroutine for a star in the yellow path. The top panel shows the $\chi^2$ distribution for the evaluated KURUCZ/ATLAS9 models across different stellar parameters ($T_{\mathrm{eff}}$, $\log g$, and [M/H]). The middle panel compares the observed spectrum (gray) with the best-fitting model (orange) for an A-type star ($T_{\mathrm{eff}} = 11\,250$\,K). The bottom panel displays the $\chi^2$ cross-correlation statistics used to determine the radial velocity of the star.}
\end{figure}

Stars with confirmed Balmer lines and $T_{\mathrm{eff}} > 15\,000$\,K are processed through the blue path using the full grid of over 46\,000 FASTWIND models, generated including only H and He features. Conversely, stars that present Balmer lines but fall below the $15\,000$\,K threshold (typically, mid- to late-B and A-type stars) are directed to the yellow path. This intermediate branch utilizes a dedicated subroutine called HiBandThereY, which relies on KURUCZ/ATLAS9 models to fully derive the stellar parameters. For these objects, the $V_{\mathrm{r}}$ computation is specifically adapted to their spectral morphology. While the main HiLineThere routine calculates a robust $V_{\mathrm{r}}$ for hot stars using narrow wavelength windows around specific features, this approach is less optimal for cooler regimes. Therefore, HiBandThereY supersedes any prior value by recomputing $V_{\mathrm{r}}$ through cross-correlation over a broad, continuous spectral range (3900--5100\,\AA). This wider region optimally exploits the full Balmer series alongside critical metallic and neutral helium lines (e.g., Si, Mg, and He\,{\sc i} $\lambda$4471). An example of a spectral fit and radial velocity determination performed by this subroutine is shown in Fig.~\ref{fig:hibandtherey}, illustrating the capability of the pipeline to automatically handle intermediate-type stars. This bridges the gap and provides reliable parameters down to 7\,500\,K, ensuring that the broadening parameters are correctly integrated for these objects.

Returning to the core HiLineThere algorithm, the routine employs an iterative procedure that simultaneously identifies spectral lines and determines a global $V_{\mathrm r}$ shift. During the iterations, selected diagnostic lines are used to calculate $V \sin\,i$ and $\zeta$. These values are then used to refine line identification and obtain a final determination of $V_{\mathrm r}$ for these early-type objects. The procedure requires at least one of the lines indicated in Table~\ref{table:lines} as a valid $V \sin\,i$ diagnostic to be present in the spectrum. These are preferably metallic lines, including Si, C, N, and Mg features. Only if none of these lines is available will a \ion{He}{i} line be used and, as the last resource, a \ion{He}{ii} line will be used, if available. In the event that no valid diagnostic line can be successfully fitted, or if the quality of the determination is insufficient, the code adopts a default value of $100\,\mathrm{km\,s^{-1}}$ for $V \sin\,i$ to ensure that the analysis pipeline can proceed without interruption. Prominent interstellar lines that may be mistaken for \ion{He}{i} lines are also searched for. The intensity of the diagnostic lines is highly sensitive to atmospheric temperature, and the tool is aware of this. During its iterative process, HiLineThere detects potential line blending or contamination that may affect the properties of the lines used for parameter diagnostics. This ensures that the influence of such effects is considered and corrected when determining stellar parameters.

\subsection{HiLineThere}\label{HLT_section}
The traditional process for classifying OB-type and RSG stars follows a straightforward approach: scientists analyze the stellar spectrum to identify characteristic absorption or emission lines. If the spectrum is dominated by H and \ion{He}{i} lines, the star is classified as early \citep{maiz2026,Gray2022}. Conversely, if these lines are absent or weak and metallic features dominate, the star is classified as a late-type. Based on these findings, the spectrum is then compared with a spectral atlas or theoretical models to determine the closest match and derive stellar parameters \citep{Walborn1991,Sota2011,Bestenlehner2025}.

Following this procedure, the first routine in our analysis workflow checks for diagnostic lines to trigger the hierarchical classification logic. The program operates iteratively, applying renormalization and contaminant removal (such as cosmic rays) before fitting a Gaussian profile to each identified line, as illustrated in Fig.~\ref{fig:WF_DETAIL}. The purpose of this initial fit is to confirm the presence of key features and determine their centers. We perform a minimum of three iterations, each refining the characterization of the spectral features to provide a robust input for the subsequent path selection and the 15\,000~K minifilter.

\subsubsection{Line fitting}\label{ssecfit}

Since the program does not initially know whether a given spectrum corresponds to an early-type or late-type star, it treats all spectra uniformly. It first attempts to identify diagnostic lines characteristic of early-type stars and derive their initial parameters ($V_{\mathrm r}$, $V \sin\, i$ and $\zeta$), since both processes must be done simultaneously. This stage serves as the primary diagnostic for the hierarchical classification logic described in Sect.~\ref{secclasif}, determining the optimal modeling path based on the detected line features. As the iterative process advances, it refines these values on the basis of the results of previous iterations, gradually achieving higher precision, all without having any prior information about the broadening parameters or $V_{\mathrm{r}}$. The fitting process is performed separately for each line, as detailed below.

\begin{itemize}
    \item The noise in the corresponding section of the spectrum is reduced through a Fourier transform, by filtering frequencies above five times the peak frequency and then transforming back to normal space. This process not only removes noise, but also weak lines considered contaminants, while preserving the main spectral components. This smoothed version is used only for identifying the line center and profile.
    \item The second derivative of the smoothed spectrum is used to identify inflection points ($d^2/dx^2 (f_{\mathrm{F-cut}}) = 0$). The closest inflection point to the theoretical line center (second column of Table \ref{table:lines}) is taken as the guess center of the line ($c_{\mathrm g}$). By evaluating the first derivative, the program also determines the concavity, that is, whether the line is in emission or absorption, by examining the sign of the derivative around the inflection point.
    \item The area ($\Delta$) between the spectrum and the locally normalized continuum is calculated to estimate the equivalent width (EW). This is done by using a simple trapezoidal rule for all the points in the selected range.
    \item Finally, the line is flagged as a candidate diagnostic line if it has a potential center and an EW above the threshold given by $ f_{\textrm{tole}}(s/n)$, which is an inverse exponential as a function of the signal-to-noise ratio, with a value of 0.01\,\AA\,as a lower limit. After this, a Gaussian profile (Eq.~\ref{eq:gaussian}) is fitted to the line. Various checks are performed to verify the quality of the fit, including assessing the residuals between the fit and the model. If the initial Gaussian fit fails to meet the criteria, a double-Gaussian profile (Eq.~\ref{eq:dgauss}) is attempted to fit blended lines, and the fit is re-evaluated. In both equations, $\mu$ represents the center of the Gaussian, $\sigma$ represents its width, and $A$ represents the amplitude. The fitting method used in all cases is based on finding the least squares by iterating over each function variable. This fit provides an estimate of the depth of the line and its possible center.
\end{itemize}

The equations used for the fits are as follows:

\begin{equation}\label{eq:gaussian}
G(x,\mu,\sigma,A) =A \frac{1}{{\sigma \sqrt {2\pi } }}e^{{{ - \left( {x - \mu } \right)^2 } \mathord{\left/ {\vphantom {{ - \left( {x - \mu } \right)^2 } {2\sigma ^2 }}} \right.
\kern-\nulldelimiterspace} {2\sigma ^2 }}},
\end{equation}

\begin{equation}
\begin{split}\label{eq:dgauss}
    DG(x,\mu_1,\mu_2,\sigma_1,\sigma_2,A_1,A_2) &= A_1 \frac{1}{{\sigma_1 \sqrt {2\pi } }}e^{{{ - \left( {x - \mu_1 } \right)^2 } \mathord{\left/ {\vphantom {{ - \left( {x - \mu_1 } \right)^2 } {2\sigma_1 ^2 }}} \right.
\kern-\nulldelimiterspace} {2\sigma_1 ^2 }}} \\ & \quad + A_2 \frac{1}{{\sigma_2 \sqrt {2\pi } }}e^{{{ - \left( {x - \mu_2 } \right)^2 } \mathord{\left/ {\vphantom {{ - \left( {x - \mu_2 } \right)^2 } {2\sigma_2 ^2 }}} \right.
\kern-\nulldelimiterspace} {2\sigma_2 ^2 }}}.
\end{split}
\end{equation}

The parameters obtained for the Gaussian profile, such as width, amplitude, and zero point (the base of the Gaussian), are stored and used to provide additional information about the line in subsequent iterations. We utilize the {\tt fmin} function from the {\tt scipy.optimize} Python library. The initial parameters for an optimal approximation are as follows. The width is determined according to the following equation: 

\begin{equation}
\label{eq:width}
    \sigma = \sqrt{\left|\sum\left(\frac{(X-x)^2(F-1)}{\sum{(F-1)}}\right)\right|} \:,
\end{equation}

where \ensuremath{F} represents the flux normalized, $x$ corresponds to the center and $X$ is the array of points. The amplitude is estimated on the basis of the absolute maximum of the line, adjusted to zero, and multiplied by the sign of the area according to the following equation:
\begin{equation}
\label{eq:amp}
    A = \max{|F-1|}sign\left(\sum{(F-1)\Delta X}\right),\end{equation}

where $\Delta X$ is the wavelength step between consecutive points. Retaining this sign assigns the proper positive or negative value to the amplitude, ensuring the correct classification of the line as either absorption or emission.

\begin{figure}
\begin{center}
\resizebox{\hsize}{!}{\includegraphics{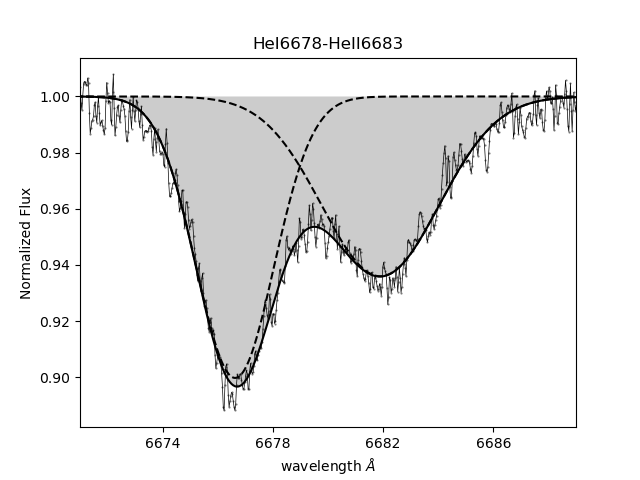}}
\caption{\label{fig:dgauss}Results obtained from fitting a double Gaussian function. This process is triggered when the residuals between the single Gaussian fit and the spectrum exceed a 3-$\sigma$ limit above the noise. The observed spectrum is shown in gray. The simple Gaussians are represented as dashed lines, while the double Gaussian is shown as a black solid line. The area corresponding to the $\Delta$ value is highlighted in gray.} 
\end{center}
\end{figure}
    
In cases where two lines blend, whether by broadening, resolution, or intrinsic properties of those lines within certain temperature ranges, it becomes crucial to account for these blended features to ensure accurate parameter determination. One notable example, which occurs in most O-type stars, is the double line composed by \ion{He}{i}~6678\,\AA\ and \ion{He}{ii}~6682\,\AA, where the absorption strength of one line decreases while the other increases, depending on effective temperature. As shown in Fig.~\ref{fig:dgauss}, a double Gaussian function is used to estimate the correct parameters for each line and determine which component is dominant, allowing for a correct identification of the center of the lines.

\subsubsection{Line selection criteria} \label{lineselectcriteria}

\begin{figure}
\begin{center}
    \resizebox{\hsize}{!}{\includegraphics{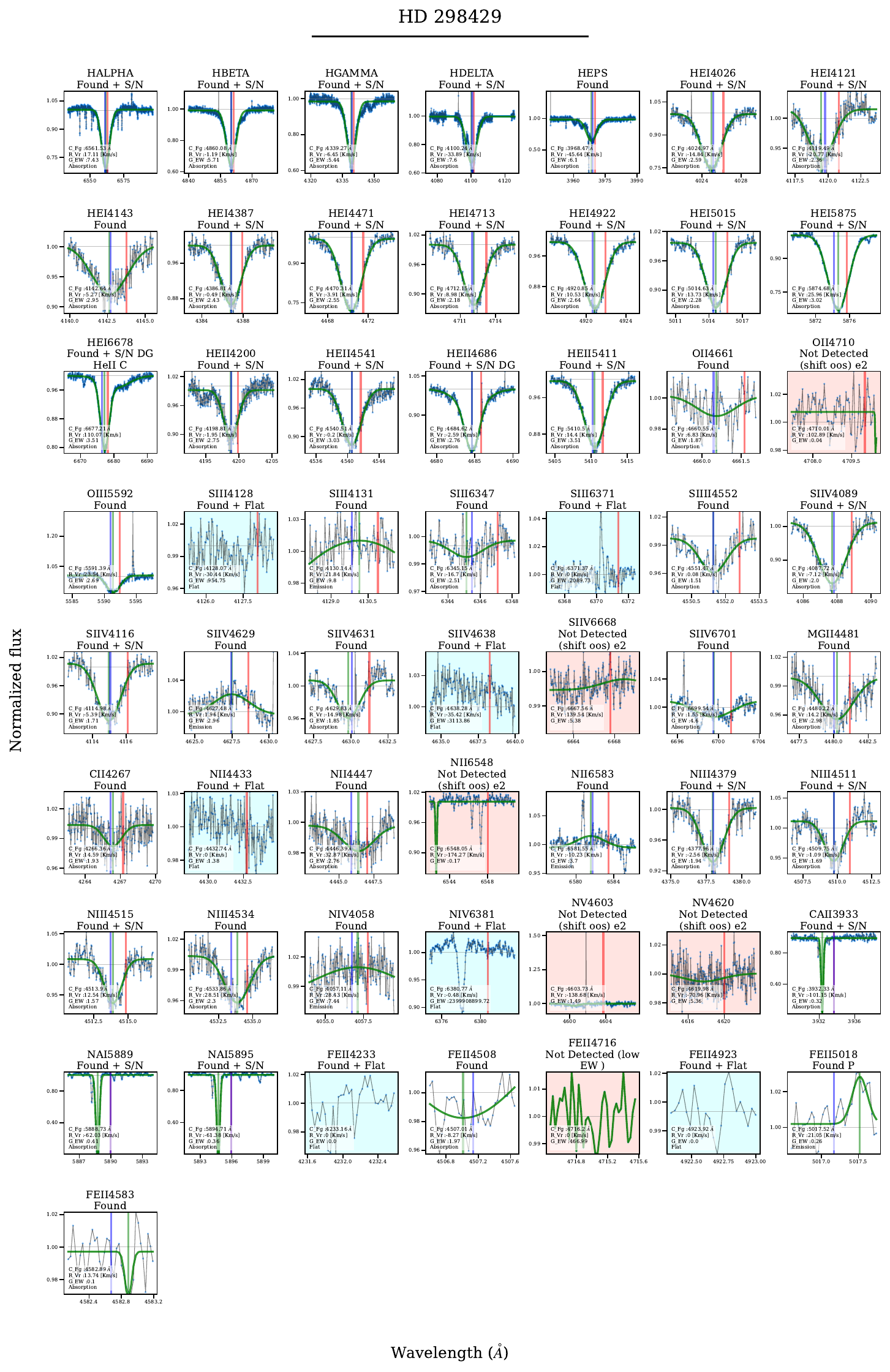}}
    \caption{\label{fig:profile}Output of the classification interface. The lines detected are shown with the label "Found"  with or without the "S/N" label, along with the best Gaussian or double Gaussian fit in green. The laboratory center, uncorrected for $V_{\mathrm r}$, is shown by a vertical red line, while the corrected center is indicated by the blue vertical line, and the Gaussian center from the best fit is marked by a green vertical line. When the profile does not fit a Gaussian but passes the filter as flat, it is displayed in a cyan background. Lines labeled as "Non-detection," according to the selection criteria, are shown in a light red background.} 
\end{center}
\end{figure}

Using the fitted line profile of each diagnostic line, we apply a scoring system to evaluate their parameters and assess the reliability of line detection. These scores provide the quantitative basis for the hierarchical classification and the path-selection logic. Detection criteria are described below.
\begin{itemize}

\item Detection threshold. If the EW is greater than $f_{\textrm{tole}}$, the line is considered detected. This is illustrated in Fig.~\ref{fig:profile}.

\item Line classification. Once detected, the line is classified as absorption or emission on the basis of the sign of the EW.

\item Displacement tolerance. We have defined a displacement ratio related to the resolution as $\delta_{\textrm{rv}}=c/R$, where $c$ is the speed of light and $R$ is the resolution. In this sense, a detected line is retained if the displacement between the center found and the center corrected by the $V_{\mathrm r}$ of the star does not exceed $50\:$km\:s$^{-1}$ plus $\delta_{\textrm{rv}}$ (see next subsection). In the first iteration, as no $V_{\mathrm r}$ has yet been calculated for the spectral lines, this criterion is not applied, but a maximum limit for displacement is set at $450\:$km\,s$^{-1}$.

\item Quality criteria. A detected line is considered a quality detection if it strictly meets the five thresholds and conditions summarized in Table~\ref{tab:quality_criteria}. Condition IV is specifically implemented for lines affected by narrow nebular emission located at the core of the line, a frequent occurrence in the spectra of young massive stars embedded in H\,{\sc ii} regions. Meanwhile, Condition V accounts for cases where the cleaning process did not successfully eliminate cosmic rays or the spectrum corresponds to an emission-line star (such as Be or WR stars), which are not analyzed in this version of the software.
\end{itemize}

\begin{table}
\caption{Quality detection criteria for spectral lines.}
\label{tab:quality_criteria}
\centering
\begin{tabular}{lp{0.8\columnwidth}}
\hline\hline
Cond. & Description / Threshold \\
\hline
I   & Absolute value of Gaussian amplitude $\geq 2\sigma_{\mathrm{range}}$ \\
II  & Residual (observed $-$ fit) $\leq 0.5\sigma_{\mathrm{range}}$ \\
III & Profile not completely flat; meaningful Gaussian fit ($\sigma > 10^{-3}$) \\
IV  & No point closer to line center than Gaussian $\sigma$ deviates by $> 3\sigma_{\mathrm{fit}}$ \\
V   & Flux amplitude $\leq 3 \times$ normalized flux \\
\hline
\end{tabular}
\tablefoot{
Lines with a flat profile (failing Cond. III) are still included to indicate ion absence.
}
\end{table}

\subsubsection{Iterations}\label{seciter}

The iteration process can be summarized as follows: The first iteration blindly searches for the lines listed in Table~\ref{table:lines}, identifying diagnostic lines that may be present, and uses them to make an initial estimation of $V_{\mathrm r}$ and line broadening. This first guess is needed to properly align the spectrum in the next iterations and to guide the subsequent classification process. The default value for $V_{\mathrm r}$ is set to 0 km\:s$^{-1}$ and the search extends up to a limit of $\pm 450$  km\:s$^{-1}$ (7.5\,\AA\ estimated at 5000\,\AA). The second iteration repeats this process with the position of the lines shifted to correct for the initial $V_{\mathrm r}$. This correction permits better line identification and the potential detection of additional lines. At the end of this iteration, a new value of $V_{\mathrm r}$ is calculated, based on all detected lines. In the third and following iterations, the routine locates the diagnostic lines from which the broadening parameters, $V \sin\, i$ and $\zeta$, will be calculated.

The Fourier technique \citep{Deeming1975,Gray1975} is first applied to isolate the pure rotational broadening component $V \sin\, i$, which is then held constant. Subsequently, the macroturbulence function is varied to achieve the best fit to the observed spectral line profile. The default values during the first iteration are 100\:km\:s$^{-1}$ for $V \sin\, i$ and 0.1\:km\:s$^{-1}$ for $\zeta$. The broadening parameters both due to stellar rotation or macroturbulence are critical for comparing theoretical models, which are generally computed without these broadening, with observed spectra.

The observed spectrum $\ensuremath{D}(\lambda)$ is represented as the convolution of several broadening factors with the original flux of the star $\mathscr{F}(\lambda)$, as indicated by 

\begin{equation}\label{eq:observed}
    {D}(\lambda)=\mathscr{F}(\lambda)*G(\lambda)*M(\lambda)*I(\lambda) \: .
\end{equation}

The simplest of these factors is instrumental resolution $\ensuremath{I}(\lambda)$, which is modeled as a Gaussian function. The second is the stellar rotation $\ensuremath{G}(\lambda)$, represented by $V \sin\, i$, which is particularly significant for early-type stars and less so for late stars. Finally, the macroturbulence $\ensuremath{M}(\lambda)$ , denoted as $\zeta$, also plays a key role for early-type stars \citep{ssimon2014,ssimon2014b}. All three parameters, as well as $V_{\mathrm r}$, necessary to align the observed lines with their rest position, can be derived from the diagnostic lines present in the spectrum, which are listed in Table~\ref{table:lines}.

The method used to determine $V \sin\, i$ is based on the Fourier transform of the line profile to generate a power spectrum. From this power spectrum, the first zero of the derivative is identified. This zero corresponds to the star's pure rotational velocity under the assumption that the line profile can be accurately modeled with a Fourier-Bessel function \citep{Deeming1975}, referenced as \ensuremath{G}  in Eq.~\ref{eq:Bessel}. This technique follows the theoretical framework proposed by \citet{Carroll1928, Carroll1933a, Carroll1933b} and \citet{Dravins1990} and, more recently, \citet{Reiners2003} and \citet{ssimon2006, ssimon2007}:

\begin{equation}
\label{eq:Bessel}
G = \frac{2(1-\epsilon)\sqrt{y} + \frac{\pi \epsilon y}{2}}{\pi\delta \left(1 - \frac{\epsilon}{3}\right)} 
\,\,\,\,\,\,\,\,\,\,\,\, 
\text{with}\,\,
\begin{array}{l}
y = 1 - \left(\frac{\lambda_0}{\Delta \lambda}\right)^2 \\
\epsilon = 0.6 \\
\delta = \lambda_0 \cdot \frac{V \sin i}{c}
\end{array}.
\end{equation}

To determine $\zeta$, we used the estimation introduced by \citet{Gray1975}, which models the macroturbulence as a radial-tangential equation, both contributing to line broadening. The function $\Theta(v,\theta)$ represents the macroturbulence broadening for a local region, as shown in Eq.~\ref{eq:Rtan}. In this context, $A_{\mathrm{T}}$ and $A_{\mathrm{R}}$ denote the fractional areas of the surface, while $\zeta_{\mathrm{R}} $ and $\zeta_{\mathrm{T}}$ represent the contributions of macroturbulence in the radial and tangential directions, respectively:

\begin{equation}
\label{eq:Rtan}
 \Theta(v,\theta)=\frac{A_{\mathrm R}}{\pi^{1/2}\zeta_R\cos\theta} 
 e^{\left( \frac{-v^2}{(\zeta_R\cos\theta)^2} \right)}
 +\frac{A_{\mathrm T}}{\pi^{1/2}\zeta_{\mathrm T}\sin\theta} 
 e^{\left( \frac{-v^2}{(\zeta_{\mathrm T}\sin\theta)^2} \right)}.
\end{equation}

Following previous authors \citep{ssimon2013,Takeda2017}, we use $A_{\mathrm R}$ = $A_{\mathrm{T}}$ and $\zeta_{\mathrm R}  = \zeta_{\mathrm T} =\zeta_{\mathrm{RT}}$, and thus the equation transforms into the sum of a simple isotropic Gaussian function and a $\delta$ function, that is,

\begin{equation}
\label{eq:RtanS}
\Theta(v)=\frac{A_{\mathrm{RT}}}{\pi^{1/2}\zeta_{\mathrm{RT}}} \left( e^{\frac{-v^2}{\zeta_{\mathrm{RT}}^2} } +\frac{-v\pi^{1/2}}{\zeta_{\mathrm{RT}}}\delta\left(\frac{-v^2}{\zeta_{\mathrm{RT}}^2} -1 \right)   \right ).
\end{equation}

\begin{figure}[ht]
\begin{center}
\resizebox{\hsize}{!}{\includegraphics{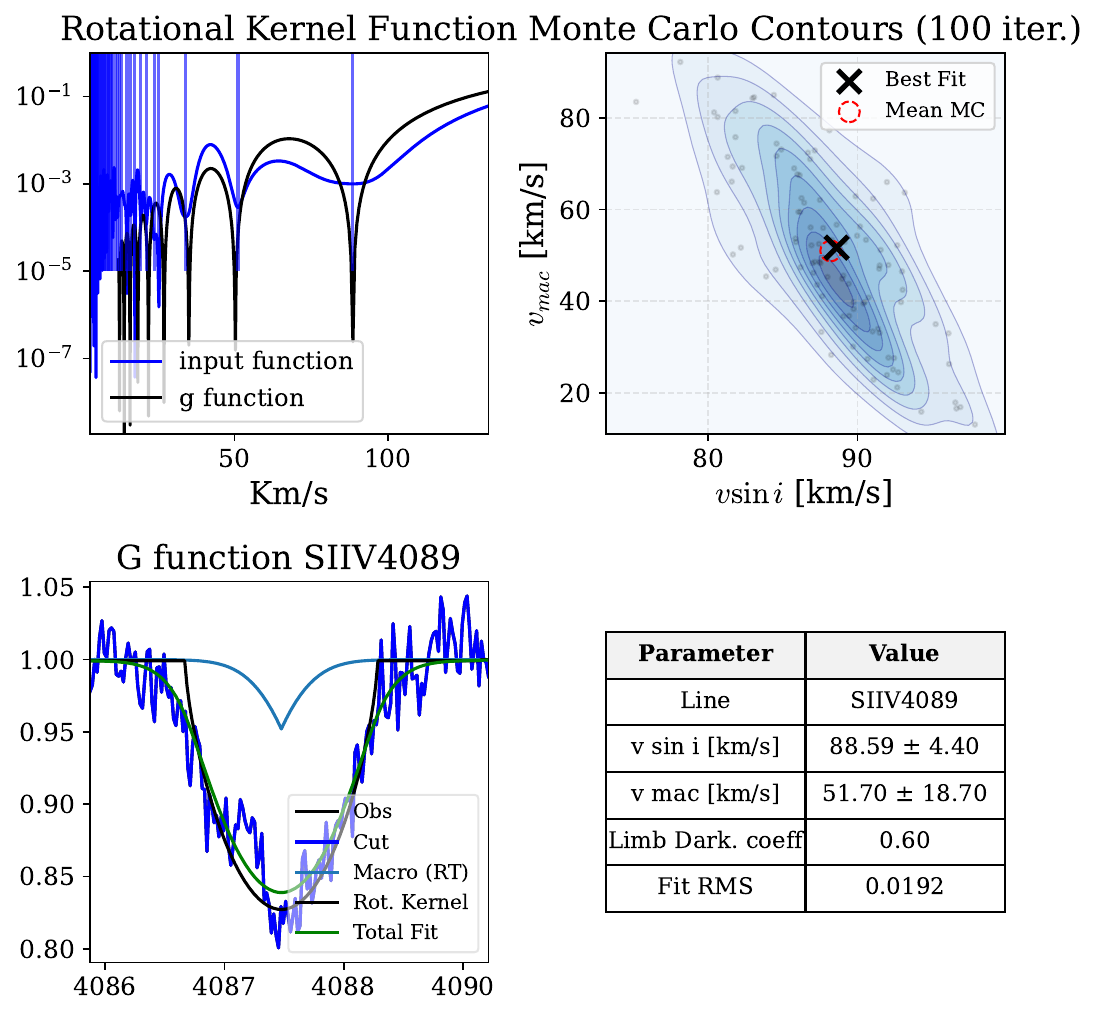}}
\caption{\label{fig:vsini} Example output for determining the projected rotational velocity and macroturbulence components using the \ion{Si}{iv}~4089\,\AA\ line. The top-left panel illustrates the determination of $V \sin\, i$ by identifying the first zero in the Fourier transform (blue line), alongside the analytical Bessel function (black line). The top-right panel displays the contours of a 100-iteration Monte Carlo simulation used to estimate the uncertainties and map the degeneracy between $V \sin\,i$ and $\zeta$. The bottom-left panel shows the observed line profile (black) along with the individual broadening components and the final convolved model (green). Finally, the bottom-right panel summarizes the derived parameters and their corresponding $1\sigma$ error margins.} 
\end{center}
\end{figure}

The broadening parameters and $G$ are later convolved with the synthetic spectrum to create a line profile directly comparable to the observed spectrum and to obtain the necessary stellar parameters. To properly evaluate the uncertainties associated with these values, we perform a Monte Carlo simulation with 100 iterations. By injecting artificial noise, scaled to the local signal-to-noise ratio, into the line profile, this approach allows us to map the degeneracy between $V \sin\, i$ and $\zeta$, deriving reliable $1\sigma$ error bars for both parameters. The complete procedure for calculating these broadening parameters and their uncertainties is illustrated in Fig.~\ref{fig:vsini}.

It is worth noting that when one of the functions in Eq. 5 dominates, it can outweigh all the others. This is particularly evident when instrumental resolution becomes the primary broadening factor in the spectrum, which is usual at lower values of $R$. For resolving powers below $R=5\,000$, the effects of instrumental broadening become significant, and by $R=3000$, the effects of physical broadening, such as $V \sin i$ or $\Theta(v,\theta)$ can be masked, since the resolution broadening is equivalent to a rotational velocity of $100~km~s^{-1}$. Despite this, our algorithm attempts to calculate $V \sin i$ as best as possible with the available diagnostic lines, independently of the instrumental resolution. However, we have implemented specific safeguards: if no valid diagnostic lines are detected and $R < 3000$, or if the fitting process fails to converge within physical limits, the code automatically defaults $V \sin i$ to the instrumental limit ($c/R$). Furthermore, for extremely low resolutions ($R < 1000$), where the instrumental profile completely overrides any physical broadening, the algorithm bypasses the line-fitting process entirely and directly adopts the $c/R$ limit, fixing macroturbulence to a minimum default value (0.1 km s$^{-1}$) to ensure the pipeline proceeds with no error.

During the iterations of HiLineThere, the spectral range of the selected line adapts dynamically. The newly determined $V \sin\,i$ directly alters the expected width of the line and, consequently, the required analysis window. The method also permits switching to an alternative line if the current one is no longer considered a valid detection under the updated parameters. The iterative process stops as soon as the $V \sin\,i$ measurement converges, defined as a difference of less than a third of our sensitivity resolution ratio $\delta_{\textrm{rv}}$ with respect to the previous step. This threshold provides a compromise between precision and the intrinsic scatter introduced by the use of different diagnostic lines.

\subsection{Classification}\label{secclasif}

No single atmosphere model can consistently reproduce synthetic spectra across all stellar temperatures without resorting to gross simplifications (for instance, LTE). The classification process directs stars along three potential analysis workflowsâ€”the blue, yellow, or red pathâ€”guided by the hierarchical decision logic described in the previous sections.

The routine first checks for the Balmer series; their absence triggers the red path for cool-star analysis (MARCS/\textsc{SteParSyn}). If Balmer lines are present, the temperature minifilter distinguishes between the blue path (hot stars, FASTWIND) and the yellow path (intermediate-type stars, KURUCZ/ATLAS9). This intermediate branch is specifically designed to bridge the gap between 7\,500~K and 15\,000~K, ensuring that the broadening parameters ($V \sin\,i$ and $\zeta$) are correctly integrated for these often fast-rotating stars. We resort to the less elaborate KURUCZ/ATLAS9 models because B/A dwarfs and giants do not require the careful treatment given to stars with radiative winds, but also because there are currently no large model grids appropriate for A-type supergiants. In the future, we will address the analysis of this rarer spectral types.

Although the entire parameter determination workflow is completely autonomous and requires no human intervention, an inspection graph is generated upon completion of the analysis (see Fig.~\ref{fig:profile}). This plot serves purely as an a posteriori quality control tool, allowing users to visually review the lines identified by the code and the resulting classification.

\begin{figure}
\begin{center}
\resizebox{\hsize}{!}{\includegraphics{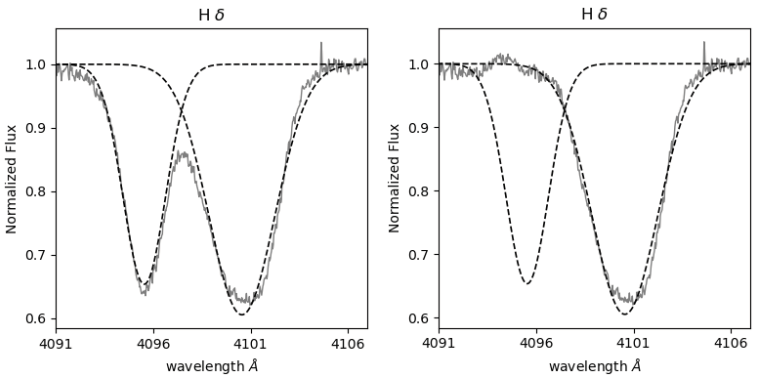}}
\caption{\label{fig:hdelta}Cleaning process for the H$\delta$ line. The left panel displays the observed line in light gray, while the two Gaussians functions fitted to the line profile are shown in dashed lines. The right panel shows the subtraction of the \ion{N}{iii} component, represented as the first gaussian, leaving a clean H$\delta$ line.}
\end{center} 
\end{figure}

In the process of cleaning spectral lines, it is essential to remove cosmic rays and other extraneous components that do not belong to the line of interest. One part of this cleaning is accomplished through a meticulous procedure that involves using the optimal model, previously derived from fitting the data, to identify and exclude regions where the residuals of the spectral line exceed a 3-$\sigma$ threshold. By applying this threshold, the method ensures that only significant features of the spectrum relevant to the subsequent analysis are considered.

Another procedure, when lines meet the minimum quality requirements, is the removal of overlapping features that contribute to the line profile. For example, in stars with spectral types around O9, there is a strong \ion{N}{iii} line potentially blending into the blue wind of H$\delta$, as illustrated in Fig.~\ref{fig:hdelta}. Similarly, other prominent lines, such as the interstellar \ion{Ca}{ii} H and K lines (at wavelengths $\lambda$ 3968 and 3934, respectively) or the \ion{Na}{i} D doublet (at wavelengths $\lambda$ 5889 and 5895), are systematically detected and cleaned. This rigorous cleaning is specifically applied to stars processed through the blue path -- either during the 1\,500-model minifilter or the final grid fitting. This ensures that the diagnostic lines that will be compared to models are free from contaminants.

\subsection{Best-fitting model, blue path}

Once a star is classified as early-type, the blue automation process is started, employing a reduced $\chi^2$ ($\chi^2_{\nu}$) minimization technique, building upon the methods described in previous works \citep[e.g.,][]{ssimon2011,ssimon2014, ssimon2014b}. This method identifies the best-fitting model from a synthetic grid of spectral models. The grid used by Astro+ was generated with the FASTWIND code \citep{puls2005, santolaya97}, which incorporates key factors such as NLTE, spherical geometry, mass loss, and line blanketing. The grid maps several parameters, including effective temperature ($T_{\mathrm{eff}}$), surface gravity ($\log\, g$), helium abundance ($\epsilon$ He), wind strength parameter ($Q$), wind velocity law exponent ($\beta$), and microturbulence (currently fixed at 10\,km\,s$^{-1}$). The wind strength parameter, $Q$, is defined as $Q =  \left[ \dot M /(R V_{\infty} )^{1.5} \right]$, combining the mass loss rate ($\dot M$), the stellar radius ($R$), and the wind terminal velocity ($V_\infty$). For this version, all stellar models were generated  with solar metallicity ($Z$/Z$_{\odot} = 1$). Other metallicities will be considered in future releases.

Although the procedure automatically computes all these astrophysical parameters, our primary focus at this stage is on accurately determining $T_{\mathrm{eff}}$ and $\log\,g$ for the observed stars. To achieve this, we have implemented a specialized subroutine to handle the H$\alpha$ line independently. H$\alpha$ is highly sensitive to stellar winds, and its profile becomes unpredictable when mass-loss parameters are not simultaneously constrained. Recombination in the extended atmosphere can fill the photospheric absorption or drive the line into emission, invalidating H$\alpha$ as a diagnostic for the static photosphere.

The subroutine detects this effect when the $\chi^2_{\nu}$ value computed for the H$\alpha$ line deviates significantly from that of the higher-order Balmer lines. Upon detection, the subroutine automatically excludes the H$\alpha$ line from the global minimization process. This prevents wind-induced profile variations from biasing the fit, ensuring an accurate derivation of the photospheric $T_{\mathrm{eff}}$ and $\log\,g$.

The extensive spectral grid, comprising 46\,000 models, was generated on the HTCondor supercomputer facility at the Instituto de AstrofÃ­sica de Canarias. This facility, which consists of a cluster of 914 cores capable of parallel processing, enabled the creation of this comprehensive grid in a few days. The models provide the line profiles for the H and He atoms, within the range of parameters outlined in Table~\ref{table:grid_parameters}.

\begin{table}[ht]
\begin{center}
\caption{\label{table:grid_parameters} Ranges of parameters used to calculate the FASTWIND grid of models.}
\resizebox{\textwidth/2}{!}{
\begin{tabular}{ l l} \hline\hline
\noalign{\smallskip}
Parameter & Range of value \\ 
\noalign{\smallskip}
\hline
\noalign{\smallskip}

$T_{\mathrm{eff}}$       & 10\,000 K to 55\,000~K (step 1\,000 K) \\
$\log\,g$       &[2.6-4.3] (step 0.1 dex)\\
Micro         &10 km\:s$^{-1}$\\
$\epsilon$He  &0.06, 0.09, 0.013, 0.017, 0.020, 0.023\\
$\log(Q)$       &-15.0, -14.0, -13.0, -12.7, -12.5, -12.3, -12.1, -11.9, -11.7\\
$\beta$       & 1.0\\
\noalign{\smallskip}
\hline \hline
\end{tabular} }
\tablefoot{This grid will be updated in further versions to also include lines from elements such as \ion{Si}{} and \ion{Mg}{}.}
\end{center}
\end{table}

\begin{figure}
\begin{center}
\resizebox{\hsize}{!}{\includegraphics{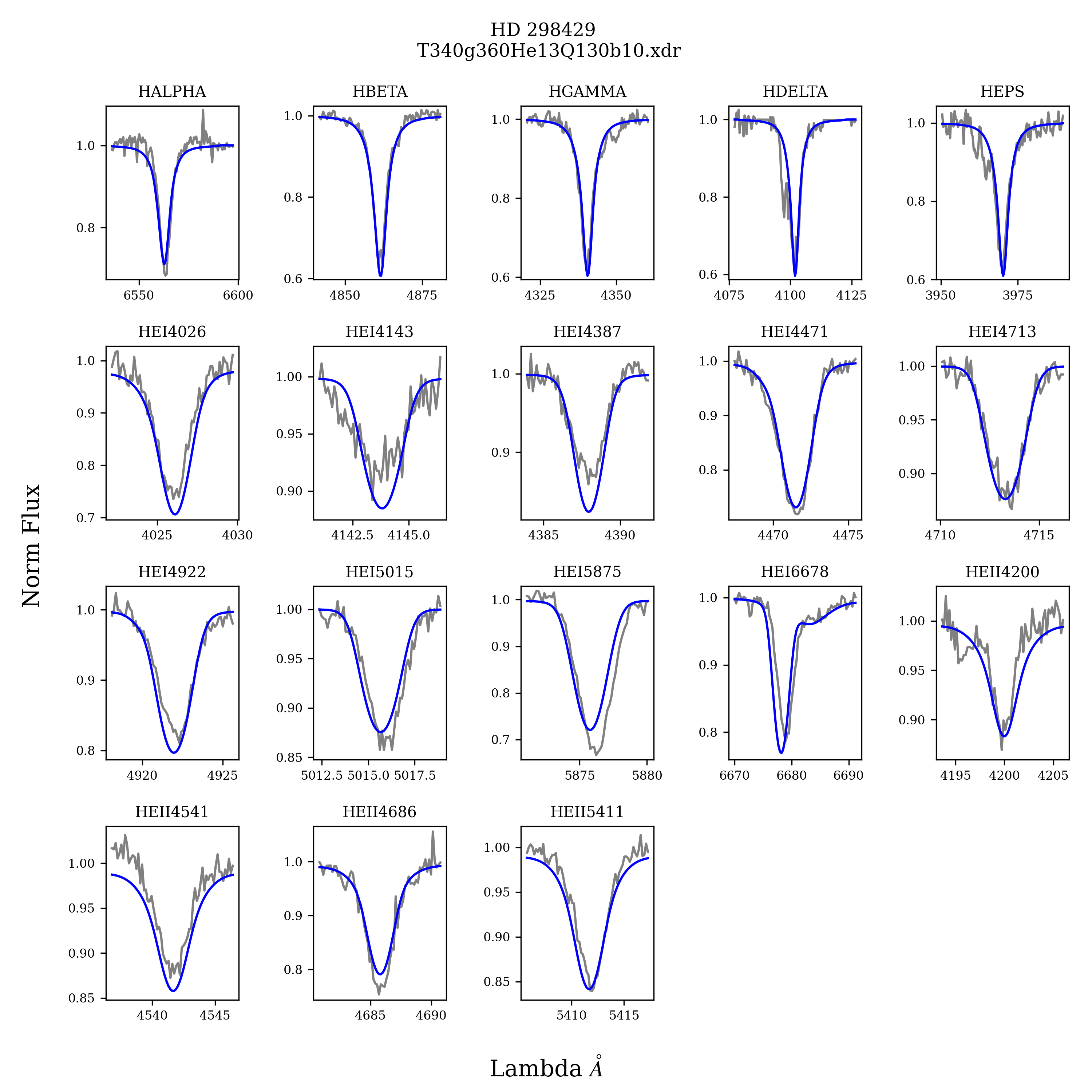}}
\caption{\label{fig:bestchi} Example of spectral fits for optical lines. Observations are shown in gray, and best-fit model profiles are in blue.} 
\end{center}
\end{figure}

The calculation of $\chi^2$ for each line requires a sequence of preparatory steps to properly align and homogenize the observed and theoretical spectra. First, the observed spectrum is corrected for $V_{\mathrm{r}}$ and the local continuum of the line is normalized, ensuring that the observed and theoretical line centers match perfectly. Next, the theoretical spectrum is modified to reproduce the observational conditions; this involves a first convolution to match the instrumental resolution, followed by a second convolution to account for rotational ($V \sin\, i$) and macroturbulent ($\zeta$) broadening. Both spectra are then resampled onto a common, equidistant wavelength grid to guarantee congruent data sets with matching data lengths. This homogenization process allows us to calculate the reduced chi-square ($\chi^2_{\nu}$) directly as follows:

\begin{equation}
    \chi^2_{\nu} = \frac{1}{\nu} \sum_{i=1}^{n} \left( \frac{O_i - E_i}{\sigma} \right)^2\: ,
    \label{eq:chi}
\end{equation}

where $O_i$ and $E_i$ are the observed and expected fluxes at each data point $i$, $\sigma$ is the estimated standard deviation (noise) of the local continuum, $n$ is the total number of data points, and $\nu$ represents the degrees of freedom. To prevent artificially inflated values in high-quality spectra, a conservative noise floor of $\sigma \ge 0.01$ (equivalent to a signal-to-noise ratio of 100) is enforced. The model that minimizes $\chi^2_{\nu}$ is then selected as the best global fit (see Fig.~\ref{fig:bestchi}).

When fitting multiple diagnostic lines, summing their individual standard $\chi^2$ values directly can introduce a severe bias: broader lines spanning more data points (such as the extended wings of the Balmer series) would disproportionately dominate the sum, overwhelming narrower, yet crucial, diagnostic lines like those of \ion{He}{i} or \ion{He}{ii}. To mitigate this effect and ensure a balanced global fit, our algorithm evaluates the number of points for each line and minimizes the global reduced $\chi^2_{\nu}$. By dividing the variance-weighted residuals by the degrees of freedom, the code prevents any single broad feature from artificially skewing the stellar parameter determination, allowing the true physical properties of the spectrum to drive the fit.

Additionally, we assign extra weight to lines that are particularly sensitive to temperature, such as \ion{He}{ii}~4541\,\AA, \ion{He}{ii}~5411\,\AA, \ion{He}{ii}~4200\,\AA, and \ion{He}{i}~4471\,\AA, as these lines, which are not strongly affected by the stellar wind, provide reliable constraints for determining the $T_{\mathrm{eff}}$ of O-type stars. In particular, the ratio \ion{He}{i}~4471\,\AA/\ion{He}{ii}~4541\,\AA\ provides the most reliable diagnostic for temperature estimation.

\begin{figure}
\begin{center}
\resizebox{\hsize}{!}{\includegraphics{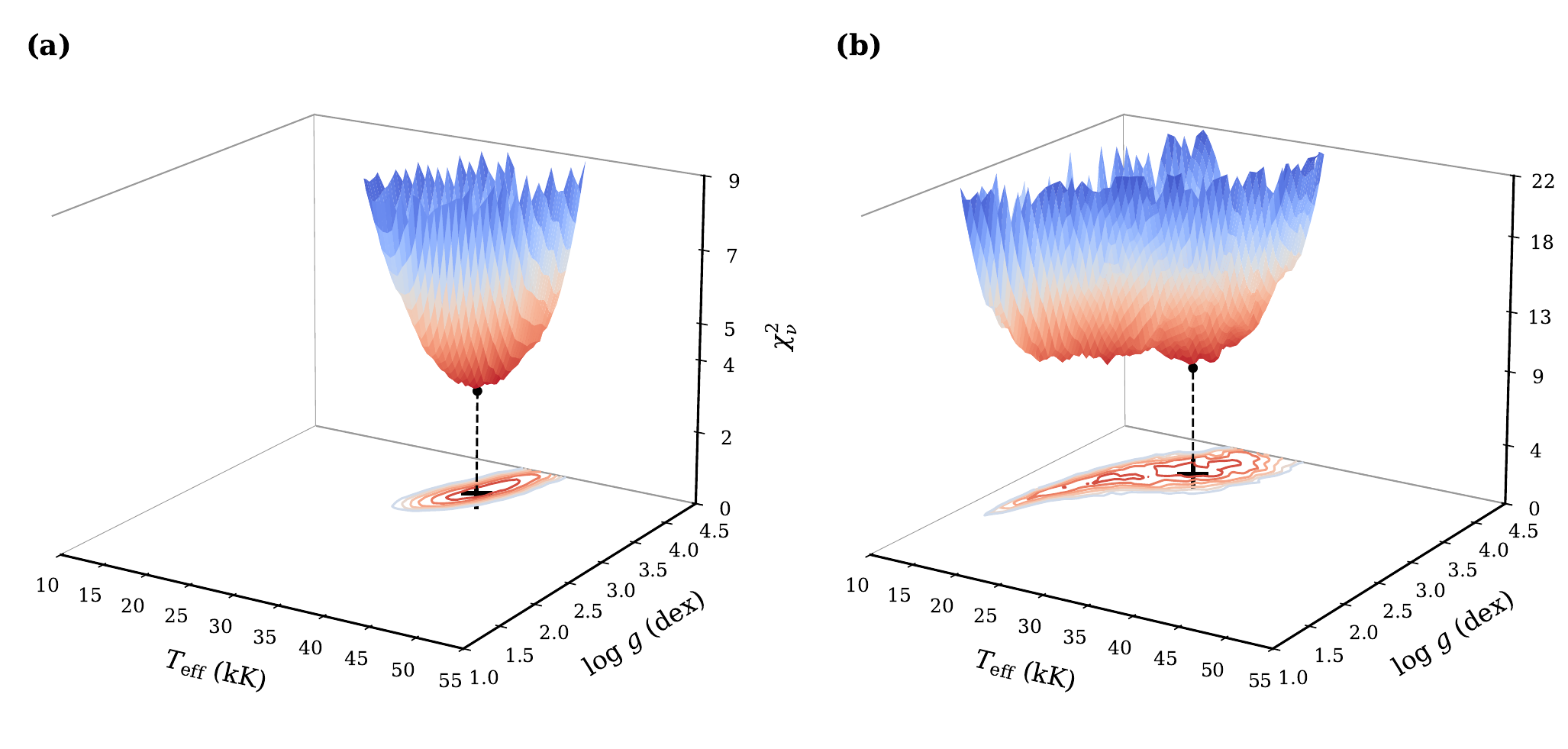}}
\caption{\label{fig:normaldeg} Reduced $\chi^2_{\nu}$ over the $T_{\mathrm{eff}}$--$\log\,g$ model grid for two example spectra, shown on identical axes. (a), left: a normal, well-determined minimum. (b), right: a degenerate case, with a broad minimum elongated along the $T_{\mathrm{eff}}$--$\log\,g$ plane. The black marker indicates the best-fitting model and the floor contours the projection of the minimum-$\chi^2_{\nu}$ envelope.}
\end{center}
\end{figure}

To obtain the final fitting value, we use the average of the area that covers one sigma of the distribution for each parameter, centered on the minimum in the $\sum \chi^2_{\nu}$ distribution, from which we determine the standard deviation at one sigma of the distribution for the error bars. It is also important to be aware that the $\chi^2_{\nu}$ distribution for the best-fit model may not always be normal, implying that it does not always present a clear minimum, such as illustrated in the left panel of Fig.~\ref{fig:normaldeg}. In some cases, the distribution may exhibit degeneracy, as shown in the right panel of Fig.~\ref{fig:normaldeg}, where the minimum is either broad and elongated along the $T_{\mathrm{eff}}$--$\log\,g$ plane or located at the edge of the grid; both situations lead to larger errors in the parameter estimation.

\subsection{Best-fitting model, red path}
\label{sec:redmodel}

If the spectrum follows the red path, the program examines its extent to determine whether it includes features that are usually considered for analysis, such as the region around the magnesium triplet or the vicinity of H$\alpha$. In their absence, the procedure focuses on the possible presence of TiO bands, which are prominent features in the spectra of cooler stars.

\begin{figure}[ht]
\centering
\resizebox{\hsize}{!}{\includegraphics{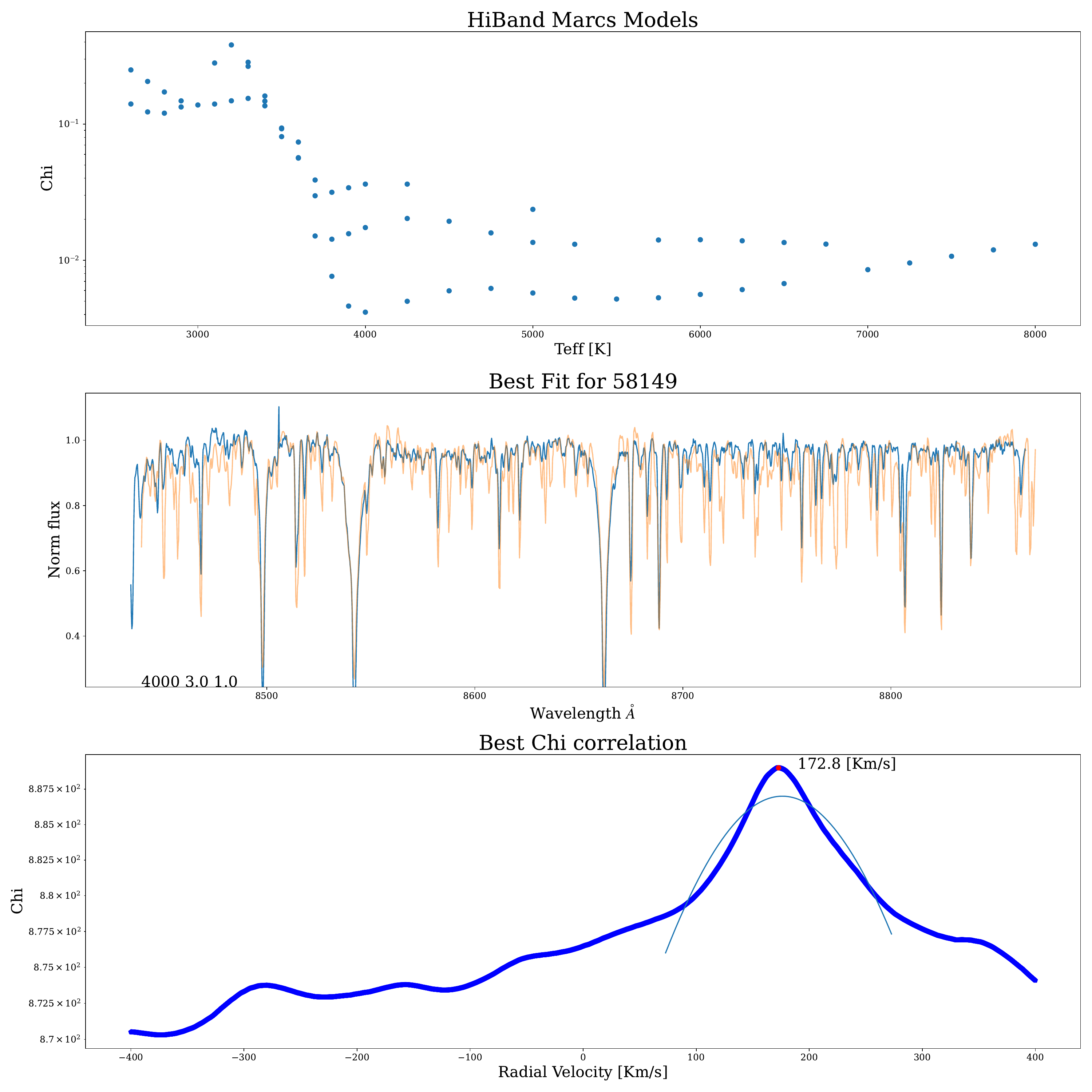}}
\caption{\label{fig:hiband} Output from the HiBandThere routine for an  RSG in the \ion{Ca}{ii} triplet (CaT) region. The top panel shows the correlation statistic distribution across the 64 MARCS model spectra considered. The middle panel compares the observed spectrum (blue) with the best-fitting model (orange). The bottom panel displays the cross-correlation over a range of $\pm400$\,km\,s$^{-1}$, alongside a Gaussian fit to the main peak (light blue curve) used to verify the symmetry and overall validity of the detected maximum.}
\end{figure}

Unlike the blue route, which uses diagnostic lines to determine $V_{\mathrm r}$, the red route takes advantage of the large number of metallic features in the spectra of cool stars and utilizes whole spectral ranges. After checking the spectral range present, $V_{\mathrm r}$ is determined by correlating a suitable section against 64 theoretical models with $T_{\mathrm{eff}}$ between 2000~K and 8000~K. This procedure, known as HiBandThere, identifies the best model by minimizing the residuals between the model and the observation. The models used in this process come from the MARCS library. The routine crosscorrRV from the PyAstronomy.pyasl package is used to find the best correlation value for $V_{\mathrm{r}}$. This procedure is illustrated in Fig.~\ref{fig:hiband}.

Once $V_{\mathrm{r}}$ is obtained, it is used for the correct execution of \textsc{SteParSyn} that provides us with the $T_{\mathrm{eff}}$, surface gravity $\log\,g$, and metallicity. \textsc{SteParSyn} is based on the use of principal component analysis (PCA) under a network of models, in this case specifically two MARCS model networks, one for $1\:\mathrm{M}_{\odot}$ stars and the second for $15\:\mathrm{M}_{\odot}$ stars. The first network encompasses temperatures from 2\,500 to 8\,000 K, gravities  $\log\,g$  between $-0.5$ and 3.0 dex, and metallicity $Z/Z_{\sun}$ between $-1.0$ and 1.0~dex. The second grid comprises temperatures from 3\,300 to 4\,500~K, gravities $\log\,g$ between $-0.5$ and 1.0~dex, and metallicities $Z/\mathrm{Z}_{\sun}$ between $-1.0$ and 0.5~dex. These two grids are used to generate interpolated models through a PCA using 70\% of the principal components.

The logic behind the use of two grids is simple, since the tool is designed to analyze spectra of massive stars. The initial study is carried out with the $1\:\mathrm{M}_{\odot}$ grid, which covers a much wider range of temperatures. If the best-fit results in temperatures below 4\,200~K, it triggers the execution of the analysis using the PCA of the $15\:\mathrm{M}_{\odot}$ models, which is more adequate for RSGs. Since this $15\:\mathrm{M}_{\odot}$ grid does not cover the entire temperature range, for stars hotter than  4\,200~K, we retain the result of the analysis with the $1\:\mathrm{M}_{\odot}$ grid. This limitation is not as important as it may seem, as the spectra of cool luminous stars are not very strongly dependent on stellar mass.

Luminous intermediate-type stars (such as A and F supergiants) that fall outside the optimal temperature bounds of these specific MARCS models are instead routed to the yellow path. In this branch, the specialized HiBandThereY subroutine takes over. Unlike its red-path counterpart (HiBandThere), which only computes $V_{\mathrm{r}}$ as an input for \textsc{SteParSyn}, HiBandThereY performs the complete parameter determination directly using the appropriate KURUCZ/ATLAS9 grids. For both MARCS grids in the red path, the analysis of the spectrum uses one of the three following spectral windows: 485\,--\,540~nm, 600\,--\,680~nm or 845\,--\,885~nm. The algorithm checks for the presence of these ranges and passes one of them to \textsc{SteParSyn}, prioritizing longer wavelengths, for the simple reason that the maximum emission for these stars is in the red and the S/N is likely to be much higher. In any event, all three ranges contain a sufficient number of metallic ion lines, such as \ion{Fe}{i}, for \textsc{SteParSyn} to provide the accurate parameters.

\section{Results and discussion} \label{ResultDiscussion}
The release of Astro+ marks the first instance of a fully automated program designed to analyze massive stars across the full range of effective temperatures. Until now, human participation was essential in optimizing the process: experts had to manually select line ranges, adjust parameters, and interpret results. Human intuition and experience were critical in ensuring the accuracy of the analysis. With the introduction of this automated system, much of this manual work is now done by the code, leading to a more efficient and consistent workflow.

Although we do not expect the results from our automated approach to match the precision and refinement of traditional methods \citep[e.g.,][]{Holgado2022,Holgado2025,Burgos2023,Martins2015,Rickard2022}, the main strength of our system lies in its ability to efficiently process large datasets in a homogeneous way. This capability allows us to assess the performance of our methods on a broader scale, even if individual results may not always provide the same level of detail as traditional techniques.

To validate the performance of Astro+, we carried out a number of tests to assess the reliability of the parameters determined. We used the O-type star sample of \citet{holgado2018}, complemented with a number of early-B-type stars from \citet{Nieva2014}, to test the blue path analysis and its transition toward cooler temperatures. To test the performance of the red path, we used the collection of RSGs from \citet{Dorda2018} to check $V_{\mathrm{r}}$ determination, along with a small sample of low-luminosity supergiants in the open cluster Berkeley~51 from \citet{Negueruela2018} to validate parameter determination. The yellow path is currently implemented as a transition module to maintain broadening parameter consistency between the hot and cool regimes. While it is fully functional and ensures algorithmic continuity across the Hertzsprung-Russell diagram, its systematic validation against a large sample of stars is deferred to future work, partly because the database currently lacks a substantial set of spectra of stars with well-determined parameters in the literature.

On the blue path, we assessed robustness with two tests: first, we utilized the same lines and weights used by \citet{holgado2018}, that is, we limited the range of spectral features employed by Astro+ for full consistency; in the second, the full spectrum with all the features considered by the code were used. On the red path, we validated the classification routine by comparing the radial velocities calculated by Astro+ with those determined by \citet{Dorda2018} in their sample. We further used the stars analyzed in \citet{Negueruela2018} to compare our fully automated implementation of \textsc{SteParSyn} with the original code.

\subsection{O-type stars}

We derive stellar parameters for the entire sample analyzed in \citet{holgado2018}, which comprises a diverse set of stellar spectra covering the full range of O-type stars, from O2 to O9.5, at all luminosity classes. This study is among the most meticulous in the literature: the sample was carefully cleaned and analyzed line by line, yielding highly detailed fits for the determination of stellar parameters. Thanks to these reliable results, we were able to identify potential systematic errors that had not been previously considered. This iterative comparison process led to progressively more complex and rigorous versions of the code, ensuring that the automated reduced $\chi^2$ minimization reproduces expert-level parameters without introducing systematic bias. As a result, the methodology remains robust and flexible, leaving the door open for future analyses involving different types of stellar spectra, such as those of B-type or cooler stars.

\begin{figure}
\begin{center}
\resizebox{\hsize}{!}{\includegraphics{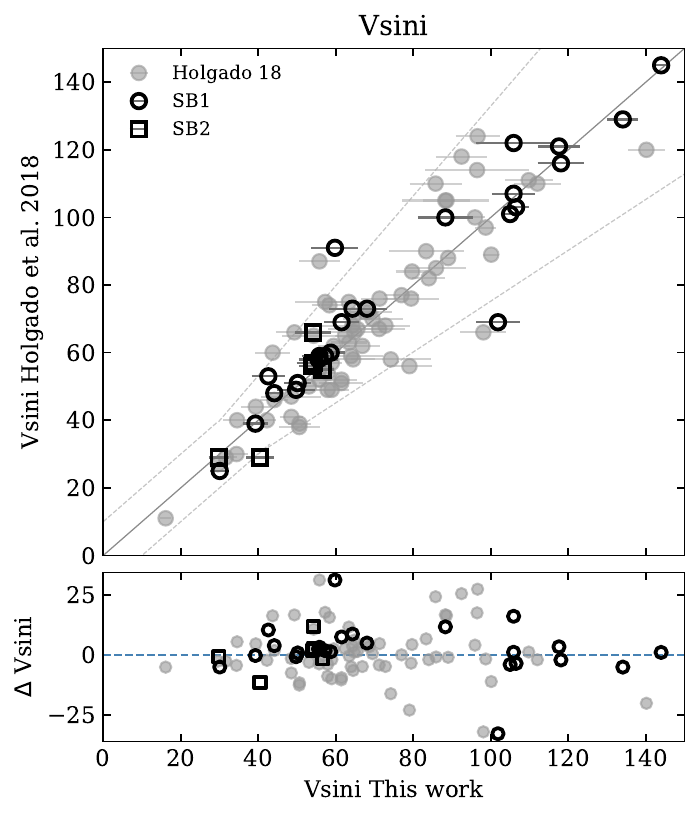}}
\caption{\label{fig:comparisonv} Comparison of the values for $V \sin\, i$ obtained by \citet{holgado2018} and by our code using the same lines. The dashed line represents $\pm 10\:\mathrm{km\,s}^{-1} $ for small values and generally $20\% $ of the value. The residual plot shows this work minus \citet{holgado2018}. Open black circles and squares mark the stars flagged as SB1 and SB2, respectively, by \citet{holgado2018}. }
\end{center} 
\end{figure}

The stellar parameters obtained by Astro+ in both tests are listed in Table~\ref{tab:listOB}. Fig.~\ref{fig:comparisonv} compares the projected rotational velocities derived by the automatic procedure against those of \citet{holgado2018}, demonstrating a strong correlation. Our methodology, which fixes the $V \sin\, i$ from the first zero of the Fourier transform before determining macroturbulence ($\zeta$), provides a more objective framework compared to manual identification. In contrast, other programs used to calculate the rotational velocity often require manual intervention to identify the position of the first zero, which can introduce subjectivity and potentially lead to incorrect identification of the rotational velocity. In our approach, after automatically selecting the first zero, we fix the value corresponding to the pure rotational profile to then determine the macroturbulent broadening function, $\zeta$ to compute and evaluate the best-fitting profile to the observed line. This allows for a more objective comparison with the theoretical models and minimizes user-dependent biases.

The comparison shows a few stars for which the values of $V \sin\, i$ are significantly different. These differences are likely justified by the different methodologies. Our choice is to always trust the result of the Fourier analysis to determine $V \sin\, i$. Contrarily, the tool used by \citet{holgado2018}, IACOB-BROAD, provides the results from both the Fourier transform and the goodness-of-fit analysis, with the latter generally used for the analysis. Despite these individual variations in the $V \sin\, i$ vs. $\zeta$ balance, the resulting total broadening profiles are nearly identical, and so the final atmospheric parameters remain unaffected.

\begin{figure}
\begin{center}
\resizebox{\hsize}{!}{\includegraphics{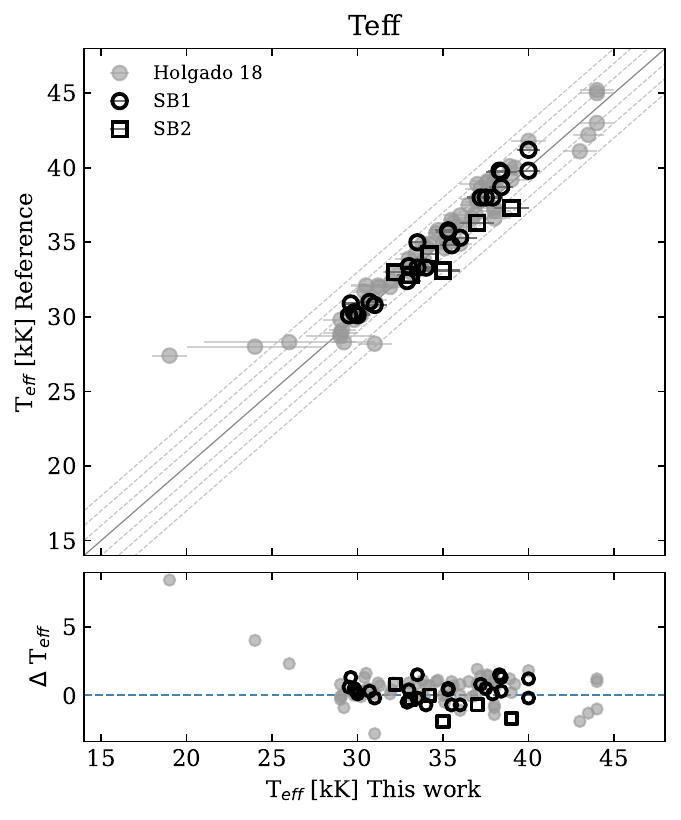}}
\caption{\label{fig:comparisont}Comparison of the values for $T_{\mathrm{eff}}$ obtained by HiLineThere in a fully automated analysis with those obtained by \citet{holgado2018}, using the same list of lines in the study. Each of the dashed lines represents $1\,000$\,K. The residual plot shows this work minus \citet{holgado2018}. Open black circles and squares mark the stars flagged as SB1 and SB2, respectively, by \citet{holgado2018}.}
\end{center}
\end{figure}

The temperature values from Astro+ also show an excellent correlation with those obtained by \citet[see Fig.~\ref{fig:comparisont}]{holgado2018}. Small differences may be attributed to variations in line range selection and the cleanliness of the line during manual selection. We note a subtle systematic trend: stars hotter than 35\,000~K tend to yield slightly lower temperatures, while those below this threshold show slightly higher values. This effect is likely linked to the automated weighting of the \ion{He}{i}/\ion{He}{ii} ionization balance and the fixed microturbulence in our current FASTWIND grid. Fitting microturbulence is not implemented in the present grid, where it is fixed at 10\,km\,s$^{-1}$; including it as a free parameter is a planned development that should help address this trend.

\begin{figure}
\begin{center}
\resizebox{\hsize}{!}{\includegraphics{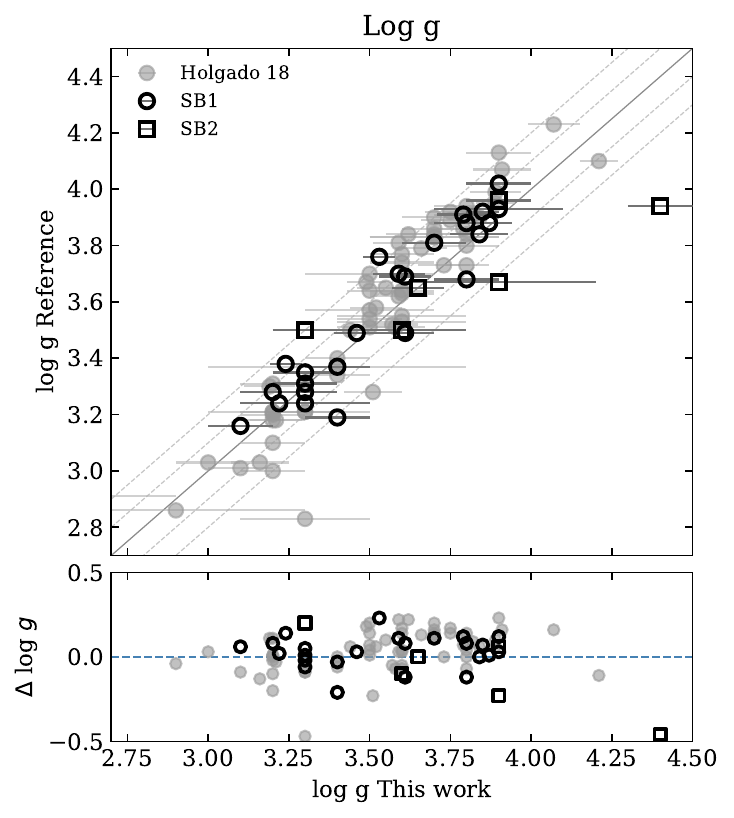}}
\caption{\label{fig:comparisong} Comparison of values for $\log\,g$ obtained by HiLineThere in a fully automated analysis with those obtained by \citet{holgado2018}, using the same list of lines; dashed lines represent $0.2$ and $0.3$\,dex. The residual plot shows this work minus \citet{holgado2018}. Open black circles and squares mark the stars flagged as SB1 and SB2, respectively, by \citet{holgado2018}.}
\end{center} 
\end{figure}

Finally, the results for $\log\,g$ (see Fig.~\ref{fig:comparisong}), while showing some scatter, remain generally satisfactory, with differences  below 0.2~dex in almost all cases and an average of 0.1~dex. As is well known, the gravity of hot stars is typically determined from the wings of the Balmer lines, and our program successfully recovers this information for a significant fraction of the stars by utilizing the reduced $\chi^2$ minimization across the H$\beta$, H$\gamma$, and H$\delta$ profiles. However, for stars with lower gravity (typically supergiants), discrepancies may arise, primarily due to the presence of additional metallic spectral lines blending into the wings of the Balmer lines. These overlapping features can introduce uncertainties in the gravity estimation, occasionally leading the automated fitting toward slightly higher gravity values to compensate for the increased equivalent width.

In the first test (Fig.~\ref{fig:comparisont}), the temperature results show an excellent correlation, except for one notable outlier at $T_{\mathrm{eff}} = 19\,\mathrm{kK}$. This object, HD~105056, is an ON-type supergiant known for its anomalous CNO surface abundances. This chemical peculiarity distorts the line profiles expected for a standard solar-metallicity grid, creating a strong degeneracy in the $\chi^2_{\nu}$ distribution. Because the code is forced to restrict its analysis to a limited, predefined subset of lines, the automated fit is driven toward a deep local minimum at cooler temperatures.

\begin{figure}
\begin{center}
\resizebox{\hsize}{!}{\includegraphics{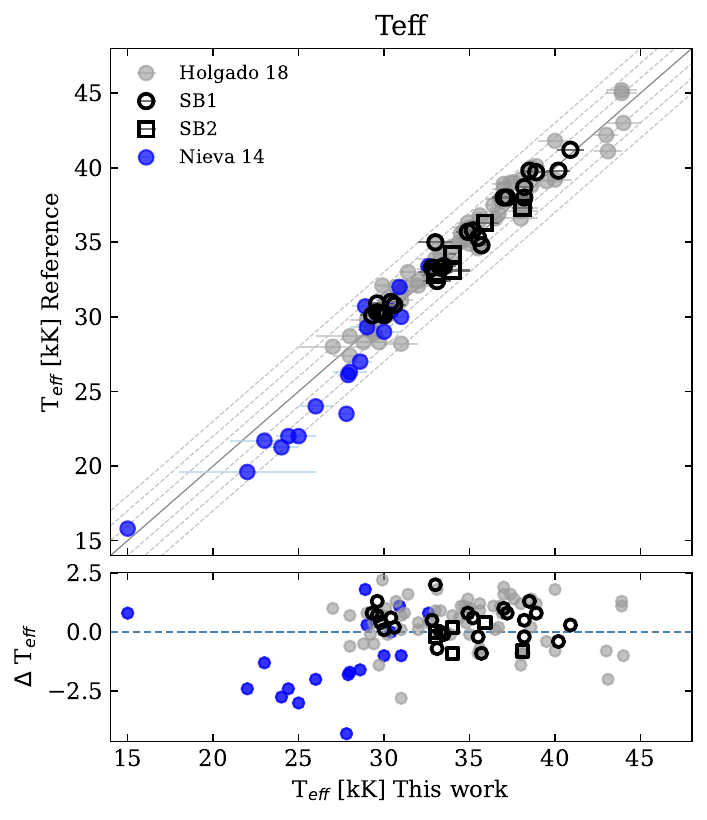}}
\caption{\label{fig:comparisont2}As in Fig.~\ref{fig:comparisont} but leaving the selection of lines free. Blue circles represent the B-type sample from \citet{Nieva2014}, highlighting the performance in the transition toward the yellow path. Open black circles and squares mark the O-type stars flagged as SB1 and SB2, respectively, by \citet{holgado2018}.}
\end{center}
\end{figure}

In the second test, we characterized the stellar spectra without prior knowledge of which lines or ranges would be available. All detected lines of \ion{H}, \ion{He}{i}, and \ion{He}{ii} were included in the analysis. This unconstrained approach fully resolves the discrepancy observed for HD~105056. By allowing Astro+ to dynamically select the diagnostic lines and apply its full weighting scheme, the inclusion of additional spectral features prevents the fit from being trapped in the false local minimum. As a result, the code correctly identifies the true global minimum at higher temperatures, matching the literature value (see Fig.~\ref{fig:comparisont2}). In principle, this approach could select broadening diagnostic lines different from those used in traditional methods. In practice, however, the automatic selection for the second test converged on the same lines as in the first test, so that the broadening parameters were unchanged.

\begin{figure}
\begin{center}
\resizebox{\hsize}{!}{\includegraphics{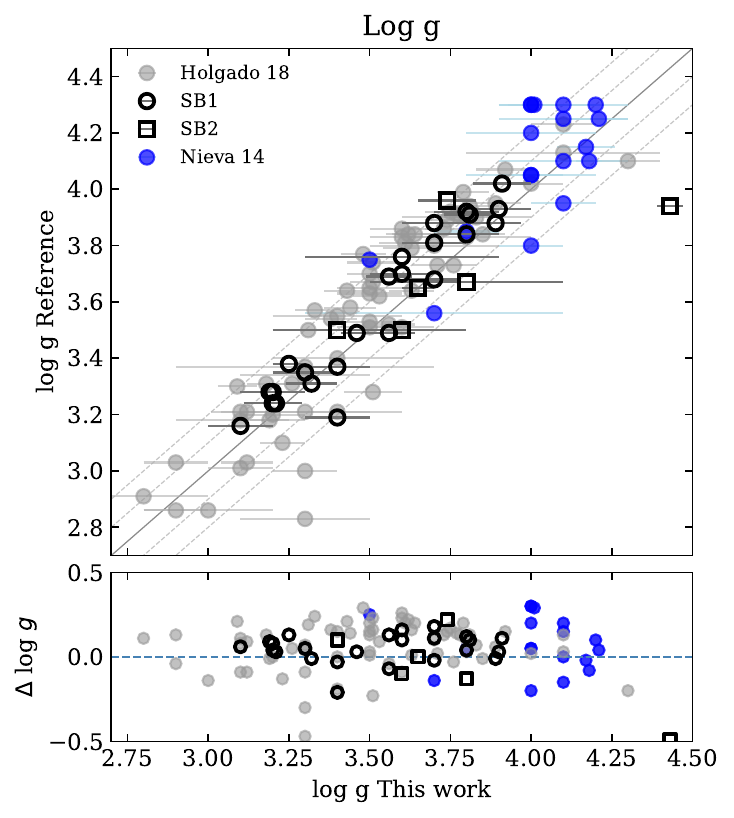}}
\caption{\label{fig:comparisong2}As in Fig.~\ref{fig:comparisong} but this time allowing the code to freely select the diagnostic lines used. Blue circles indicate the B-type sample from \citet{Nieva2014}. Open black circles and squares mark the O-type stars flagged as SB1 and SB2, respectively, by \citet{holgado2018}.}
\end{center} 
\end{figure}

With respect to stellar parameters, this second test with a broader spectral range (and therefore more lines) compared to \citet{holgado2018}, produced results that are fully consistent with those obtained using specific lines and weights, as shown in Figs.~\ref{fig:comparisont2} and~\ref{fig:comparisong2}. In fact, the differences in $T_{\mathrm{eff}}$ are now smaller, although a small systematic effect is still visible. Moreover, the inclusion of a small sample of early-B stars from \citet{Nieva2014} confirms that the automated parameter determination remains robust and consistent as the pipeline approaches the lower temperature limit of the blue path. All derived parameters are consistent with the spectral types of the stars (see Table~\ref{tab:nieva_results}). The moderate differences in $T_{\mathrm{eff}}$ with respect to the values in \citet{Nieva2014} likely stem from the use of both different input spectra and different model atmospheres\footnote{While we validate the performance for O-type stars with the same set of spectra used by \citet{holgado2018}, the early-B types are tested with different spectra of the same stars analyzed by \citet{Nieva2014}.}.  In fact, if we consider only dwarf stars, there is a very good correlation with spectral type in our values, with seamless continuity between B- and O-type stars. For $\log\,g$, the same moderate tendency toward lower values seen in the first test is still present, which can again be explained by the presence of weak lines in the wings of the Balmer lines. Despite the absence of typical diagnostic lines in some cases, the program effectively characterized the stellar spectra, determining rotational and macroturbulence velocities from alternative lines.

One key aspect of our methodology is the decision not to fix specific lines for analysis. Fixing lines can limit flexibility and introduce a bias, as it assumes that only certain lines are relevant, which may not hold true for all stars. Moreover, the algorithm is designed to process all spectra, without previous knowledge of the spectral range present, which is at odds with a preselected set of lines. By allowing the program to dynamically select the most relevant lines, we ensure that the analysis remains robust and capable of handling a variety of stellar spectra (with different spectral ranges or resolutions) without being constrained by preselected features or incomplete spectral information.

Finally, most of the differences observed between both tests can be attributed to the removal of components such as line blends, bad pixels, cosmic rays, or saturated areas. Additionally, discrepancies may arise from the selection of spectral ranges for the lines or the exclusion of certain lines from the analysis on the basis of decisions that, though informed, break the homogeneity of the analysis. Despite these factors, our fully homogeneous automated method has been shown to provide parameters that are reliable and consistent across a wide range of spectral types.

Spectroscopic binaries are common among O-type stars, and the sample of \citet{holgado2018} includes a large fraction of them: of the O-type stars compared here, 31 are flagged as spectroscopic binaries (25 SB1 and 6 SB2), a fraction consistent with the multiplicity statistics reported by those authors. These systems were kept in the comparison and are marked with open symbols in Figs.~\ref{fig:comparisonv}, \ref{fig:comparisont}, \ref{fig:comparisong}, \ref{fig:comparisont2} and~\ref{fig:comparisong2}. Their parameters are recovered with the same accuracy as those of the presumably single stars: the mean absolute differences with respect to \citet{holgado2018} are $\sim$0.6~kK in $T_{\mathrm{eff}}$, $\sim$0.1~dex in $\log\,g$, and $\sim$6~km\,s$^{-1}$ in $V \sin\,i$ for the binaries, comparable to those of the single stars ($\sim$0.9~kK, $\sim$0.1~dex, and $\sim$8~km\,s$^{-1}$). The most discrepant objects in these figures are not binaries but stars with strong wind variability, line-profile variability, or peculiar abundances, such as HD~105056. This is expected: for a single-lined system, a single-epoch spectrum is analyzed as a single star with a shifted radial velocity. Moreover, \citet{holgado2018} note that in all their double-lined systems the secondary is very faint, so that the single snapshot is fitted as an apparently single star. In these cases binarity does not bias the derived parameters but, at most, contributes to the degeneracy of the fit. For the large, automatically processed samples expected from surveys such as WEAVE, SB1 systems will be analyzed without special treatment, while SB2 systems with components of comparable brightness are expected to surface as outliers in the goodness-of-fit statistics; their identification and component-by-component analysis is a planned development.

\subsection{Late-type stars}

\begin{figure}
\begin{center}
\resizebox{\hsize}{!}{\includegraphics{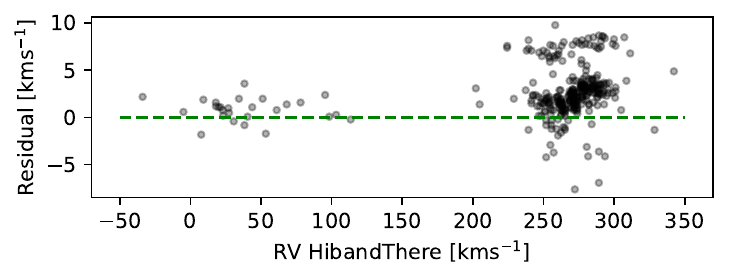}}
\caption{\label{fig:comparisonRR}Comparison between the radial velocities ($V_{\mathrm r}$ ) determined by \citet{Dorda2018} and those obtained automatically by HiBandThere. The plot shows the residuals of our results minus those of \citet{Dorda2018}.} 
\end{center}
\end{figure}

The HiBandThere routine provides the necessary parameters and input for \textsc{SteParSyn}, including the $V_{\mathrm r}$ of the star. The $V_{\mathrm{r}}$ measurements obtained by HiBandThere were compared with the values calculated by \citet{Dorda2018}, for all the stars in their 2013 campaign, which were observed in the region of the CaT at a resolution $R=10\,000$. As shown in Fig.~\ref{fig:comparisonRR}, the comparison reveals an excellent 1:1 correlation, underscoring the effectiveness of our new program. The small offset, at a level of $2$\,km\,s$^{-1}$ is a reflection of the different methodologies: \citet{Dorda2018} derive $V_{\mathrm r}$ with the method of \citet{Koposov2011}, a Bayesian comparison of the full spectrum against a set of model templates, whereas HiBandThere obtains it from a cross-correlation of the full spectral window against our model grid. Both approaches are therefore template-based; the small, one-directional offset most likely reflects differences in the model grids and spectral treatment, and remains well within the RV uncertainties reported by \citet{Dorda2018}. It is important to note a shift of approximately $7$\,km\,s$^{-1}$ that affects some objects. This subset of spectra had had a heliocentric correction applied, which corresponds to this value, while the rest were in the observatory's frame.

\begin{figure}
\begin{center}
\resizebox{\hsize}{!}{\includegraphics{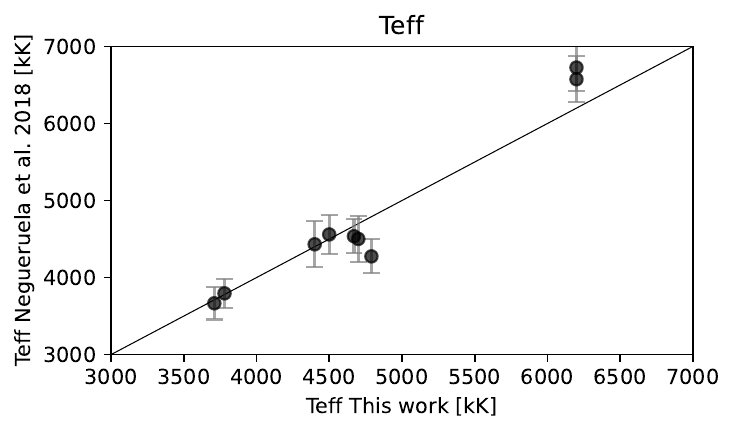}}
\caption{\label{fig:Teff_Red} Comparison of the values of effective temperature ($T_{\mathrm{eff}}$) obtained by our method using \textsc{SteParSyn} for a small sample of RSGs with those determined by  \citet{Negueruela2018} } 
\end{center}
\end{figure}

For our final validation, we determined parameters for a sample of low-luminosity supergiant stars from the open cluster Berkeley~51. This sample had previously been analyzed using the same code in \citet{Negueruela2018}. However, these authors fixed $\log\,g$  to values of 0 (for the cooler stars) or 1 (for warmer objects), whereas our grid covers $\log\,g$ values from 0 to 3. The parameters derived by Astro+ are shown in Table~\ref{listRed}. The comparison therefore focuses on the $T_{\mathrm{eff}}$ values, which are displayed in Fig.~\ref{fig:Teff_Red}. There is obviously a good correlation, but we must keep in mind that we are using the same code as in the original paper. The small discrepancies can be attributed to the differences in $\log\,g$. There is a small degree of degeneracy between $T_{\mathrm{eff}}$  and $\log\,g$, and the values derived for gravity, though poorly constrained, tend to be higher than those fixed by \citet{Negueruela2018}. Despite these minor differences, the results highlight the consistency and accuracy of the parameters obtained by the automated procedure.

\subsection{Systematic errors on blue path}

All the uncertainties considered so far result from the fitting procedure and the quality of the spectra themselves. Users should be aware that the stellar parameters provided by Astro+ are tied to the FASTWIND grid that was used to derive them. Systematic differences between the results of different stellar atmosphere codes for hot stars have been reported in the literature \citep{Massey2013, holgado2018}. In order to provide a rough estimate of such systematic errors, we performed an experiment by feeding a grid of TLUSTY models \citep{tlusty2003} into our code as if they were observed spectra.

\begin{figure}
\begin{center}
\resizebox{\hsize}{!}{\includegraphics{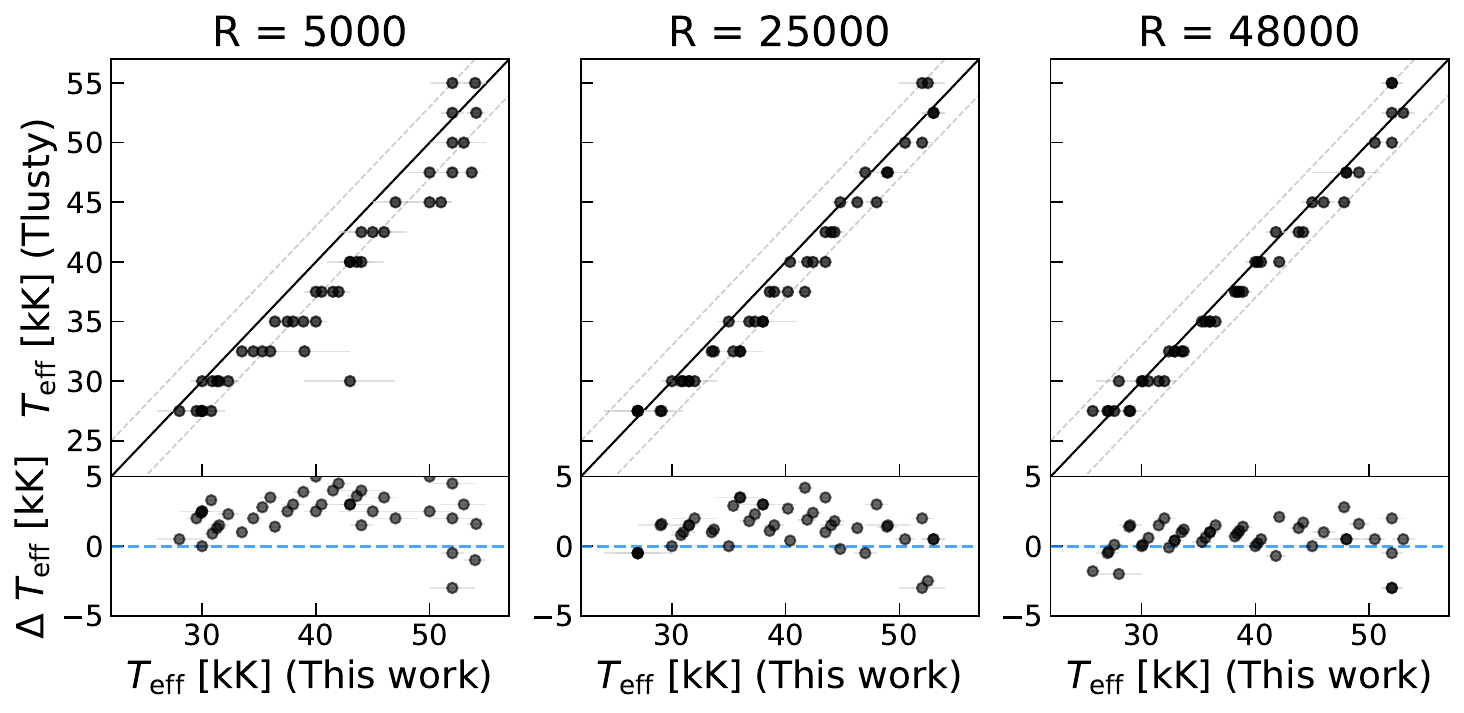}}
\caption{\label{fig:TlustyT} Comparison between the $T_{\mathrm{eff}}$ determined by our code and the label in the TLUSTY models. Dashed lines represent $\pm 2\,000$\,K. From left to right, the panels show results for resolving powers of 5\,000, 25\,000, and 48\,000. The residual plot shows the difference between the values derived by Astro+ and the catalog value for the TLUSTY models.}
\end{center}
\end{figure}

\begin{figure}
\begin{center}
\resizebox{\hsize}{!}{\includegraphics{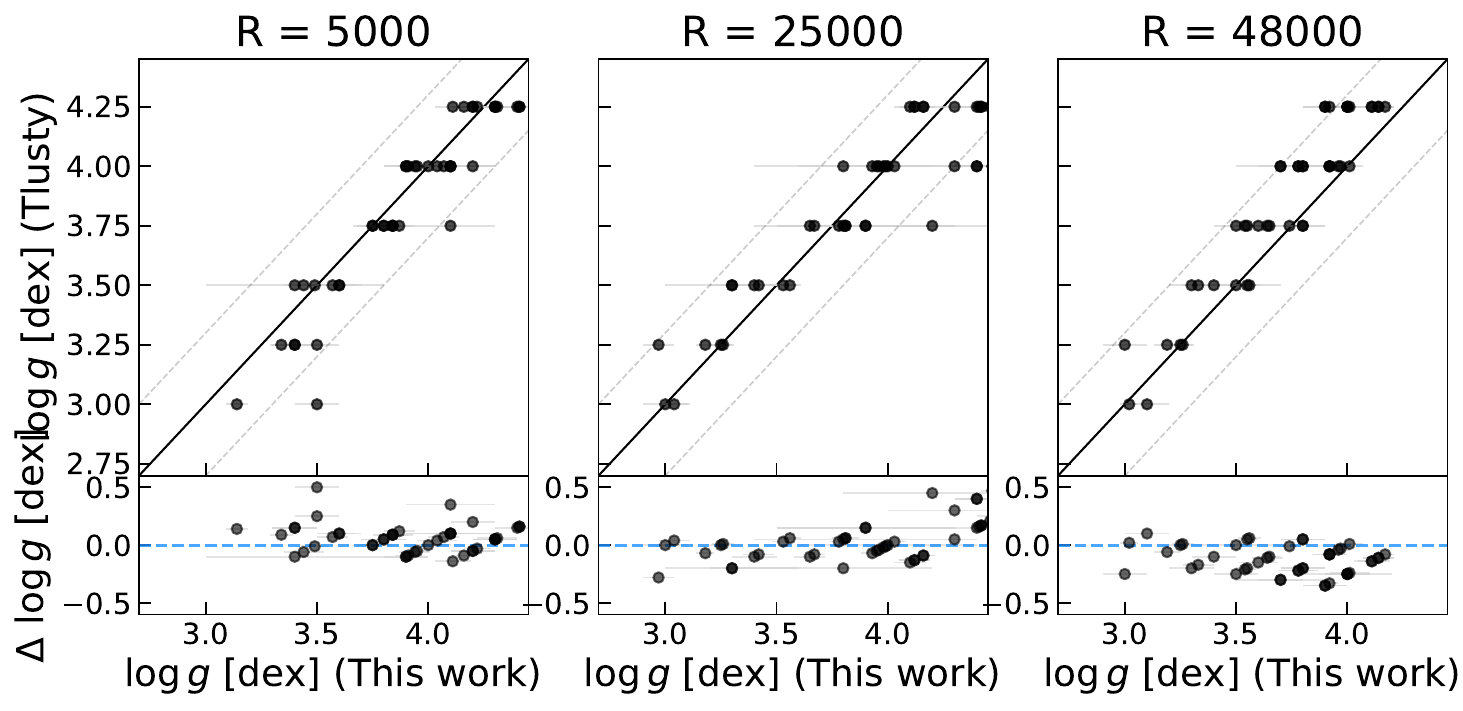}}
\caption{\label{fig:TlustyG}Comparison between the $\log\,g$ determined by our code and the label in the TLUSTY models. The panels are the same as in Fig.~\ref{fig:TlustyT}. Dashed lines represent $\pm 0.3$\,dex. The residual plot shows the difference between the values derived by Astro+ and the catalog value for the TLUSTY models.} 
\end{center}
\end{figure}

We stress that this test did not consist of a separate analysis with a different set of models. We simply uploaded the TLUSTY models as if they were input observed stellar spectra at different resolutions ($R=5\,000, 25\,000, 48\,000$), without introducing artificial noise. The TLUSTY grid used, computed at solar metallicity, covers a range of $T_{\mathrm{eff}}$ from 27\,500 to 55\,000~K and values of $\log\,g$ from 3.0 to 4.5~dex. The correlations obtained from these tests are shown in Figures~\ref{fig:TlustyT} and~\ref{fig:TlustyG}.

For $T_{\mathrm{eff}}$, the results show a good correlation with the expected errors. Only close to the edges of the grid do we see some larger discrepancies, most likely related to the differences in $\log\,g$ discussed in the following. However, for surface gravity, noticeable differences appear when $\log\,g > 3.6$~dex. This happens because the low-resolution models (shown on the left panel of Fig.~\ref{fig:TlustyG}) produce mismatches in the spectral line profiles used to determine gravity. Since this issue arises directly from the resolution applied to the TLUSTY models, the resulting scatter should not be treated as a systematic error. Additionally, it is important to note that the parameters used to convolve the line profiles are derived from the metallic lines present in the TLUSTY model. When the broadening is mainly dominated by resolution, this can lead to inaccurate rotational velocity estimates for specific models, resulting in a wrong convolution that does not properly fit the corresponding spectrum. As a result, this contributes to the larger error bars for gravity. Nevertheless, this experiment can be used to estimate an upper limit on the systematic differences implied by the use of a given set of models, which should be below $2\,000$\,K for stars in the O-type range.  We caution that this test does not compare like with like: TLUSTY is a static, plane-parallel code, whereas our grid is computed with FASTWIND, which includes a stellar wind. Part of the differences seen here therefore reflects the known systematics between different atmosphere codes, which have been explored in the literature \citep[e.g.,][]{Massey2013, holgado2018}, rather than the automated analysis itself.

\section{Conclusion}
\label{sec:conclusions}

We have developed Astro+, a comprehensive repository for storing spectra of massive stars, equipped with a robust tool for determining stellar parameters in a fully homogeneous manner, both for hot massive stars and evolved supergiants. The system is designed to fully support future releases of large spectroscopic datasets by ensuring consistent processing standards. For the analysis of early-type stars, a new fully automated tool, known as HiLineThere, has been developed. For cool luminous stars, we used a combination of a classifier HiBandThere, which also determines $V_{\mathrm{r}}$, and an implementation of \textsc{SteParSyn}.

To ensure continuity and a homogeneous treatment across the Hertzsprung-Russell diagram, Astro+ incorporates a yellow path driven by the HiBandThereY subroutine, using KURUCZ/ATLAS9 models for intermediate-type stars. While this branch bridges the gap between the hot and cool regimes, its systematic validation is deferred to future work. Our analysis pipelines have been rigorously tested using the O-type star sample from \citet{holgado2018} and a small set of early-B stars from \citet{Nieva2014} for the blue path, alongside the late-type star sample from \citet{Dorda2018} for $V_{\mathrm r}$ and the low-luminosity supergiants from \citet{Negueruela2018} for the red path. This is the first paper in a series devoted to the Astro+ database; a forthcoming paper (Paper~II) will address several of the items left for future versions, such as the systematic validation of the yellow path and the extension of the analysis to other metallicity regimes.

The results obtained by our fully automated routines compare very favorably with those determined by an expert spectroscopist, within the expected errors for a wide range of stellar parameters, but with the crucial advantage of providing a strictly homogeneous analysis across the entire dataset, free from subjective human biases. Furthermore, we used the TLUSTY model collection to obtain valuable information on the accuracy of our results and their associated error ranges.

In conclusion, the application that we have created represents a powerful and versatile tool for data analysis and visualization. With its intuitive interface and advanced capabilities, users can efficiently explore large datasets and uncover insights that would otherwise be difficult to obtain. Our programs can help advance the identification and characterization of stellar populations by providing fast, efficient, and systematically homogeneous analysis for large datasets. Using state-of-the-art algorithms and techniques, HiLineThere facilitates the efficient determination of stellar parameters for massive stars, establishing itself as an indispensable resource for researchers in the field.
\section{Data availability}

The Astro+ database, including the analyzed spectra and the derived stellar parameters, is publicly accessible at \url{https://astroplus.ua.es/}. The red-path analysis relies on the \textsc{SteParSyn} code \citep{tabernero2022}, publicly available at \url{https://github.com/hmtabernero/SteParSyn/}.

\begin{acknowledgements}
The authors acknowledge the financial support of the Spanish Ministerio de Ciencia e InnovaciÃ³n (MCIN) with funding from the European Union NextGenerationEU and Generalitat Valenciana (GVA) in the call Programa de Planes Complementarios de I+D+i (PRTR 2022), HIAMAS project (reference ASFAE/2022/017). Further support has been provided by the GVA under grant PROMETEO/2019/041, and the MCIN and Agencia Estatal de InvestigaciÃ³n under grants PID2021-122397NB-C21/C22 and PID2024-159329NB-C21/C22 (MCIN/AEI/10.13039/5011\,00011033/FEDER, UE). S.~S.-D. acknowledges support from the European Commission (EC) under the project OCEANS -- Overcoming challenges in the evolution and nature of massive stars, HORIZON-MSCA-2023-SE-01, grant agreement No.~101183150. Funded by the European Union. L.~R.~P. acknowledges support by grants PID2022-140483NB-C22 and PID2022-137779OB-C41 funded by MCIN/AEI/10.13039/501100011033 and by ``ERDF A way of making Europe.'' H.~T. acknowledges support from Spanish grants PID2021-125627OB-C31 and PID2024-158486OB-C31 funded by MCIU/AEI/10.13039/501100011033 and by ``ERDF A way of making Europe,'' by the programme Unidad de Excelencia Mar\'ia de Maeztu CEX2020-001058-M financed by MCIN/AEI/10.13039/501100011033 and by the MaX-CSIC Excellence Award MaX4-SOMMA-ICE, by the Generalitat de Catalunya/CERCA programme, and by the European Research Council (ERC) under the European Union's Horizon Europe programme (ERC Advanced Grant SPOTLESS; no.~101140786). Views and opinions expressed are however those of the author(s) only and do not necessarily reflect those of the European Union or the European Research Council. Neither the European Union nor the granting authority can be held responsible for them.
\end{acknowledgements}

\bibliographystyle{aa}
\bibliography{biblio}

\begin{appendix}
\nolinenumbers

\section{The Astro+ database: guidelines for its usage}\label{sec:database}

The following sections describe these functionalities in detail.

\subsection{Ingestion}

Data obtained from telescopes can be stored in various formats, including flat American Standard Code for Information Interchange (ASCII) files and the widely used Flexible Image Transport System (FITS) format. However, this variety can create challenges when reviewing and comparing different observations, as different methods of encapsulating information can be applied, even within the same observatory, depending on pipeline versions.

To address this issue, we have developed a web interface that provides a well-organized, structured platform where users can upload one or several stellar spectra. The system supports both FITS files, with configurable settings, or simple ASCII flat files, which must be accompanied by a header CSV file for metadata correlation. Upon upload, data are first stored on our temporary servers and saved as files, while being integrated with our SQL database (MySQL), which manages file uploads and various workflows. Our versatile and user-friendly interface allows for the simultaneous upload of multiple spectra, with a batch limit of 500 Mbytes to prevent server bottlenecks. Compressed files, including gz, tar, tgz, zip, and rar formats, are supported, regardless of the spectrum format type.

The correct interpretation of the contents within the files, such as header data, primary spectrum information, wavelength, and flux, is essential to the interaction between the user and the database. Once the user confirms that the spectrum is correctly displayed and identified, it can be submitted to an administrator for final preservation in the permanent database.

For ASCII files, the upload process is straightforward, but requires an additional header file to ensure proper metadata correlation. For FITS files, we offer 15 different combinations for extracting wavelength and flux data (e.g., deriving wavelength by using CRVAL/CDELT or extracting both wavelength and flux as arrays). 

\begin{figure}[ht]
\begin{center}
\includegraphics[width=0.45\textwidth]{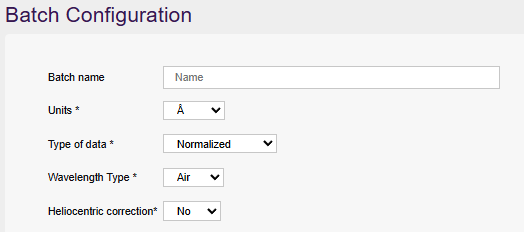}
\caption{Screenshot of the Astro+ webpage illustrating the basic requirements for the upload. These include providing the unit of the wavelength, the data type, and the possibility of naming the batch.}
\label{fits0}
\end{center}
\end{figure}

The file upload process starts by prompting the user to provide specific information (see Fig.~\ref{fits0}), including an optional batch title (if none is provided, the batch number is used), the wavelength unit (with $\AA$ as default), and whether the spectrum is normalized, non-normalized, or flux-calibrated (default). Additionally, users must specify whether the wavelength is referenced in vacuum or air (default: air), and whether the wavelength is heliocentrically corrected (default: no correction). Once this information is provided, the user can proceed with file selection in ASCII or FITS format.

\subsubsection{FITS}

\begin{figure}[ht]
\begin{center}
\includegraphics[width=0.45\textwidth]{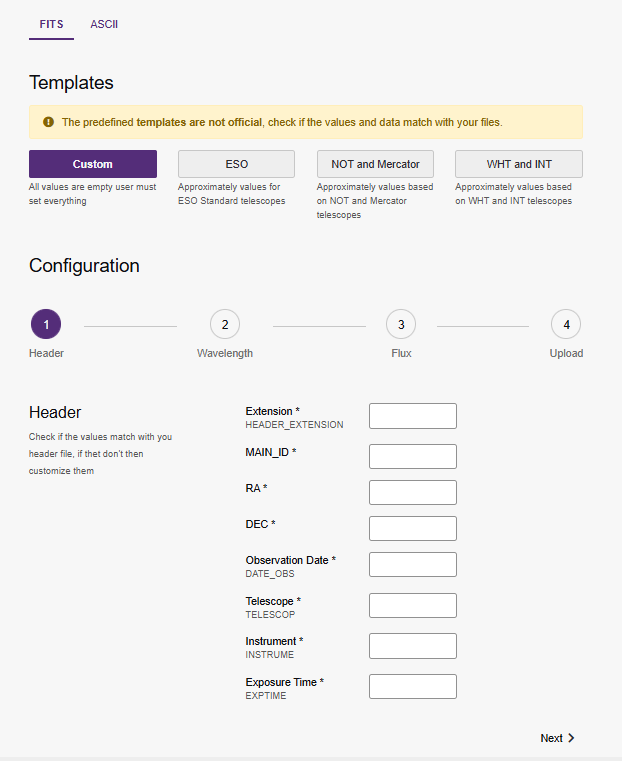}
\caption{\label{fits1}Screenshot of the Astro+ interface illustrating the first steps of uploading the FITS file, showing the mandatory header fields.}
\end{center}
\end{figure}

Since FITS files can have various configurations, we have predefined three of the most commonly used configurations based on representative telescopes, such as the Very Large Telescope (VLT), Nordic Optical Telescope (NOT), and the William Herschel Telescope (WHT)(see Fig.~\ref{fits1}). The user can modify these settings as needed, by selecting the option labeled as "CUSTOM." The details of these predefined templates are described in the following paragraphs.

\begin{figure}[ht]
\begin{center}
\includegraphics[width=0.45\textwidth]{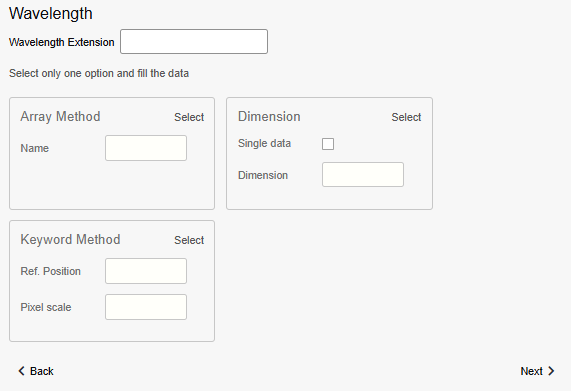}
\caption{\label{fits2}Screenshot of the Astro+ webpage  illustrating the second step of uploading the FITS file, the wavelength configuration. The array, dimensioning, and keyword methods are shown.}
\end{center}
\end{figure}

\begin{figure}[ht]
\begin{center}
\includegraphics[width=0.45\textwidth]{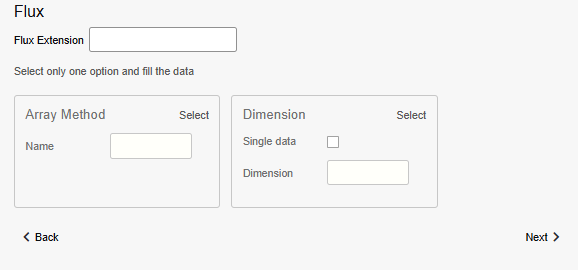}
\caption{\label{fits3}Screenshot of the Astro+ webpage  illustrating the third step of uploading the FITS file, the flux data.}
\end{center}
\end{figure}

\begin{figure}[ht]
\begin{center}
\includegraphics[width=0.45\textwidth]{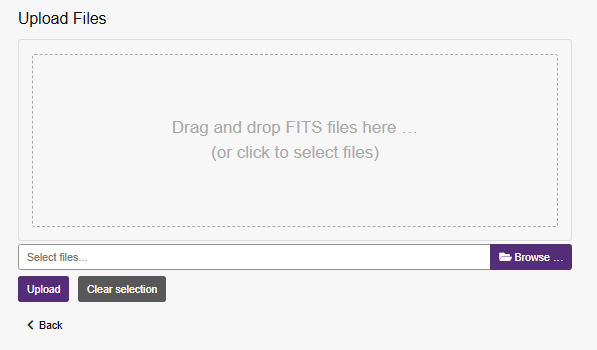}
\caption{\label{fits4}Screenshot of the Astro+  webpage illustrating the fourth  step of uploading FITS file, the loading of the files}
\end{center}
\end{figure}

The configuration of headers and data is divided into four steps:

\begin{itemize}
    \item[1]: The first step requires the user to specify the extension (HEADER\_EXTENSION) where the headers are located and provide the keywords that identify mandatory data for extraction, including the object name (MAIN\_ID), right ascension (RA), declination (DEC), observation date (DATE\_OBS), telescope and instrument (TELESCOPE, INSTRUME), and exposure time (EXPTIME).

    \item[2]: Wavelength: As illustrated in Fig.~\ref{fits2}, the user must specify how to extract the wavelength from the FITS file. Three methods are available: array (when the wavelength is stored as an array in the binary data), dimension (if the wavelength is embedded in one of the binary data dimensions), or keyword (if the wavelength is calculated by using CRVAL and CDELT1). For example, the VLT template uses the array method with the WAVE keyword, while NOT and WHT use the keyword method with CRVAL1 and CDELT1.

    \item[3]: Flux: As shown in Fig.~\ref{fits3}, the user must specify how to extract flux data. Similarly to the wavelength configuration, the user can choose the array method (if the flux is stored as an array in the file extensions) or the dimension method (if the flux is embedded in one of the binary data dimensions). For example, the VLT template uses the array method with the FLUX keyword, while the NOT and WHT use the dimension method, extracting flux from dimension 0.

    \item[4]: Upload: As shown in Fig.~\ref{fits4}, the user can select files individually, in batches, or as compressed archives containing multiple FITS files for simultaneous upload.
\end{itemize}

\subsubsection{ASCII}
\begin{figure}[ht]
\begin{center}
\includegraphics[width=0.40\textwidth]{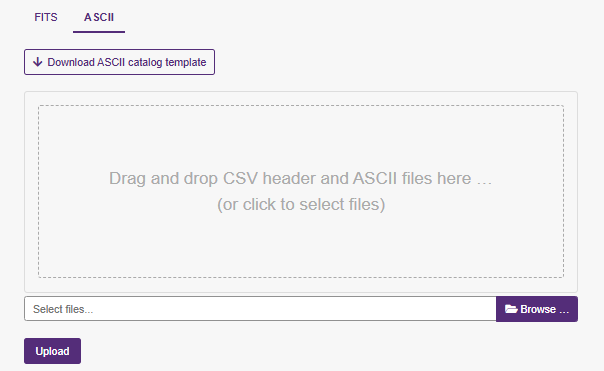}
\caption{\label{ascii}Screenshot of the Astro+ webpage  illustrating the only step involving uploading ASCII files}
\end{center}
\end{figure}

The file upload process for ASCII files is more straightforward but requires additional preparation (see Fig.~\ref{ascii}). Since ASCII files typically lack information about the source or acquisition of the spectrum, users must provide a CSV file containing the name of the ASCII file along with its corresponding headers. This CSV file, along with the ASCII data file(s) for one or multiple spectra, should be uploaded to the designated window. The CSV file must include the fields FILENAME, MAIN\_ID, DATE\_OBS, RA, DEC, TELESCOP, INSTRUME, and EXPTIME, which need to be filled in for each filename.

\subsection{Inspection}
\begin{figure}[ht]
\begin{center}
\includegraphics[width=0.45\textwidth]{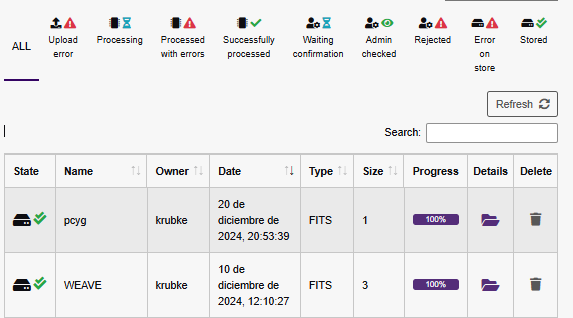}
\caption{\label{batcha}Screenshot of the Astro+ webpage  illustrating the batch screen. At the top of the list is the legend of all the available statuses in the upload process. }
\end{center}
\end{figure}

\begin{figure}[ht]
\begin{center}
\includegraphics[width=0.45\textwidth]{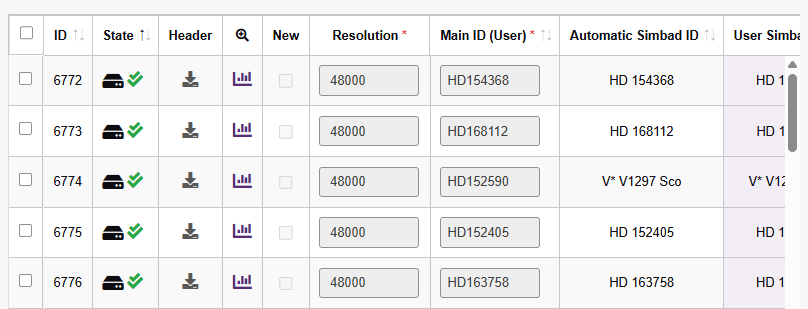}
\caption{\label{batchb}Screenshot of the Astro+ webpage illustrating the batch detail screen. In the image, shows a list of stars that compose the batch. In this interface, the user must verify the star names and add the resolution for each spectrum. If all the stellar spectra have the same resolution, the top box allows the user to apply the same resolution to the entire batch, which can then be saved.}
\end{center}
\end{figure}

\begin{figure}[ht]
\begin{center}
\includegraphics[width=0.45\textwidth]{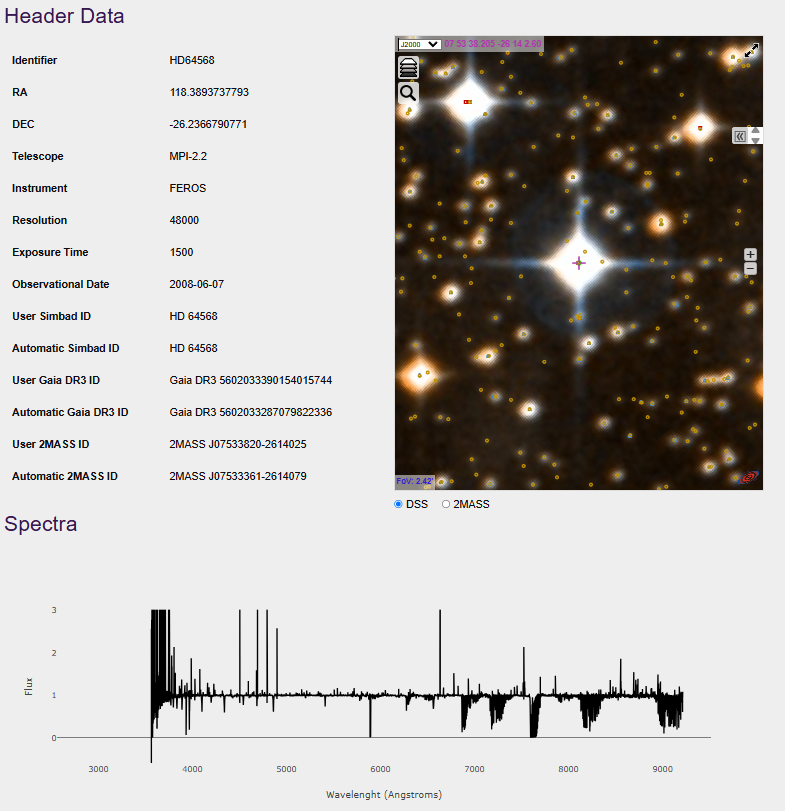}
\caption{\label{batchc}Screenshot of the Astro+ webpage  illustrating the spectrum detail. The top left is the header of the file containing the spectrum. The top right is the API Aladin, which confirms the object's position. The bottom panel plots the data, allowing the user to check if it was read correctly.}
\end{center}
\end{figure}

The inspection hierarchy follows a structured process: A batch consists of multiple spectra, and each spectrum has its own set of parameters. It is the user's responsibility to verify that the batch has been correctly uploaded at each step. As shown in Fig.~\ref{batcha}, the user can review their uploaded batches and their upload status. By selecting a specific batch, the user gains access to the list of stellar spectra included (see Fig.~\ref{batchb}). At this stage, each uploaded spectrum is converted into Astro+'s standard FITS format, and the star's information is cross-referenced with the Gaia, 2MASS, and SIMBAD databases through their respective APIs. The user must confirm the star names and IDs and then input the resolution for each spectrum. In cases where a star is not found in any of these public catalogs, the user has the option to designate it as a newly cataloged star. By clicking on a star within the batch, the user accesses a detailed view of the spectrum uploaded together with the reference parameters retrieved from APIs cross-referencing. This is the first opportunity for the user to visually inspect the spectrum and verify that it has been correctly processed. The user can also verify that the correct star has been assigned to the spectrum, by using the Aladin API to check the field of view (see Fig.~\ref{batchc}).


\begin{figure}[ht]
\begin{center}
\includegraphics[width=0.4\textwidth]{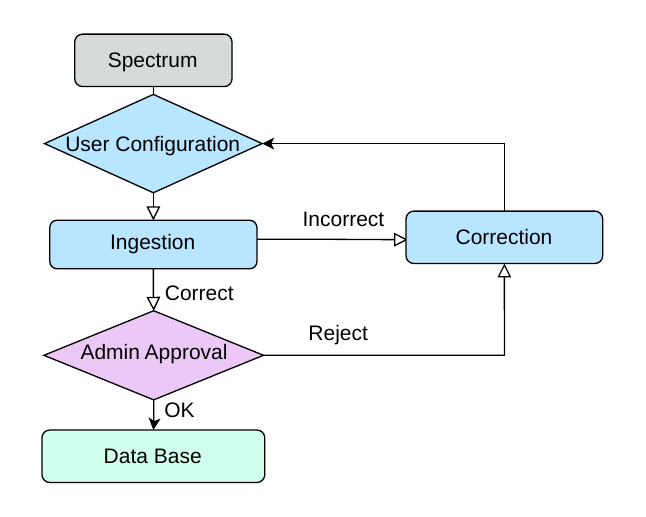}
\caption{\label{fig:WF_Ingest}
Complete upload workflow. The diagram details each step, from the initial data selection to the final confirmation of a successful upload, including the interactions between the user and the administrator during the process.} 
\end{center}
\end{figure}

In the final step, within the batch details window, the user submits the verified batch to the administrator for final review and integration into the permanent database. The role of the Astro+ administrator is to verify that the spectrum is correctly loaded and configured, for example, checking that the wavelength is accurate, the resolution matches the instrument, and the star name was correctly obtained from the API databases or properly configured as a newly cataloged star. If any spectrum requires corrections, the administrator returns the batch to the respective user for modification. Once approved, the star will be accessible in the permanent database search engine. The complete workflow is illustrated in Fig.~\ref{fig:WF_Ingest}.

\subsection{Public data}
\label{sec:data}

\begin{figure}[ht]
\begin{center}
\includegraphics[width=0.45\textwidth]{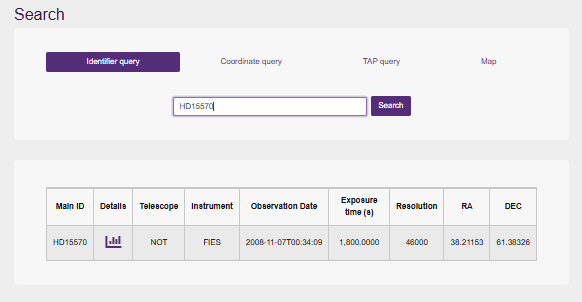}
\caption{\label{fig:Search}Screenshot of the Astro+ webpage demonstrating one of the search options available in our database. In this example, the search output displays information for the star HD 23302.} 
\end{center}
\end{figure}

One of the main services of our web application is to provide the scientific community with tools to search, share, and inspect stellar spectra, and obtain derived stellar parameters from the uploaded spectra. Users can search for stars in our database using one of these three methods: by star ID (see Fig.~\ref{fig:Search}), by stellar coordinates within a defined search cone, and through a more advanced query via SQL, which allows filtering by various derived stellar parameters. These parameters include ($T_{\mathrm{eff}}$), surface gravity ($\log\,g$), rotational velocity ($V \sin\, i$), or macroturbulence ($\zeta$) (for early-type stars), or metallicity and $V_\mathrm{r}$ (for late-type stars), if available. The query results generate a list of stars with spectra stored in our database that match the search criteria. Users can inspect each spectrum individually, with the option to download it if needed. A field recording the provider of each spectrum and its primary reference, so that other users can identify the original source and cite it alongside the Astro+ paper, is planned for an upcoming release.

\onecolumn

\section{Astro+ in detail}

\begin{figure}[!htb]
\begin{center}
\includegraphics[width=0.7\textwidth]{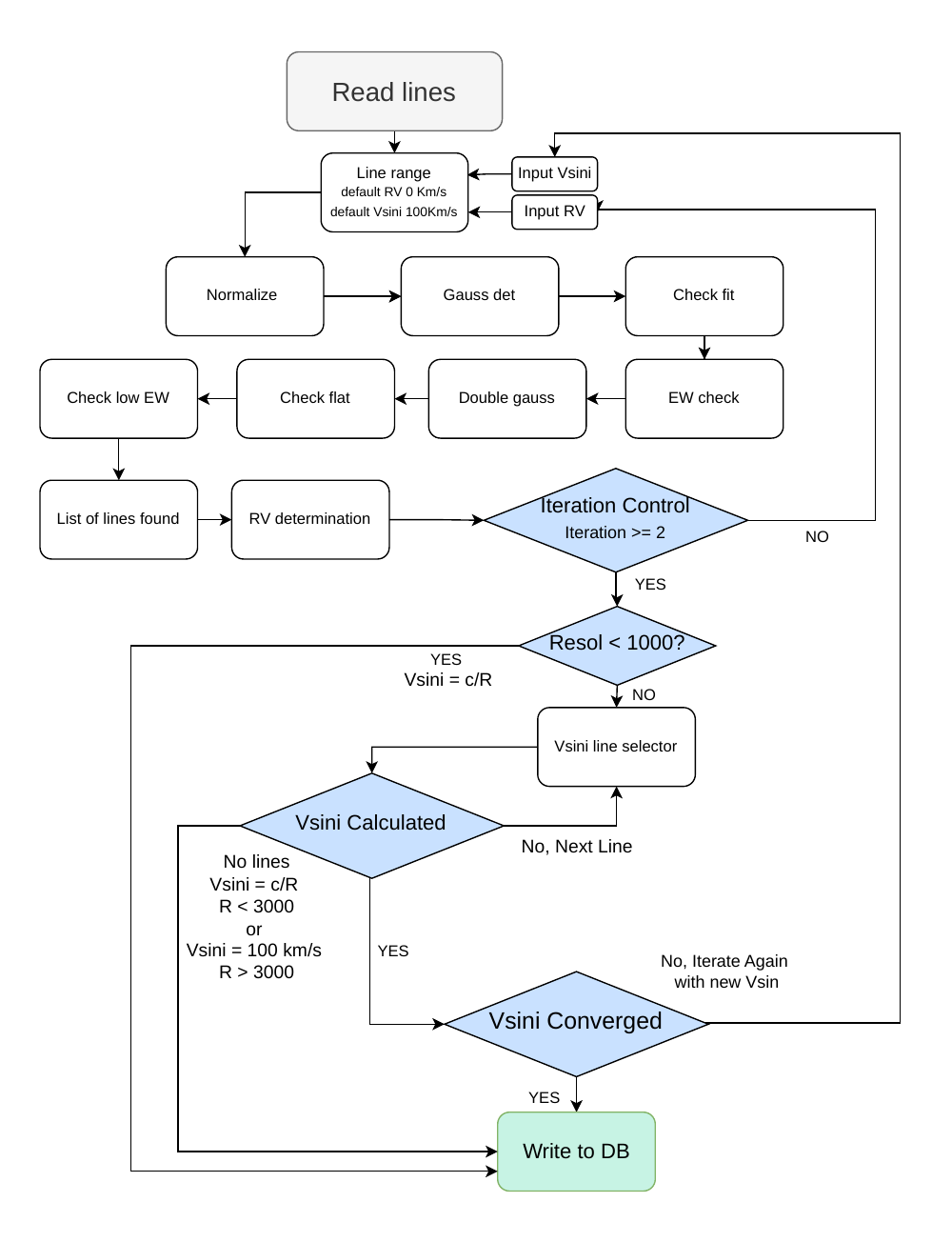}
\caption{\label{fig:WF_DETAIL}
HiLineThere workflow for spectrum classification. The workflow follows a series of steps to ensure accurate analysis. First, line ranges are defined based on the radial velocity and rotation parameters, followed by local normalization of the continuum. A Gaussian fitting function is then applied to determine the line profile, and the residuals are analyzed to evaluate the quality of the fit. The equivalent width (EW) is estimated and compared to the result of the Gaussian fit. If double components are detected, a double Gaussian function is tested for potential improvement. If a Gaussian fit cannot be applied, line flatness is checked, especially for absent lines. The reliability of the Gaussian fit is further verified for high signal-to-noise lines with low EW. Identified lines are summarized, and $V_{\mathrm r}$ is calculated. This process is repeated to refine the estimations, until $V \sin\, i$ and $\zeta$ are determined, if possible, by using one of the lines listed in Table~\ref{table:lines}. The iteration control incorporates specific safeguards to bypass the fitting process or adopt theoretical limits ($c/R$) for extremely low-resolution spectra ($R<1000$) or when no valid lines are found. The process ends when the $V \sin\, i$ converges, as explained in Section~\ref{seciter}, or when there are no lines left in the list to obtain an estimation. Finally, the derived parameters are written to the database, providing the necessary input to route the spectrum toward the blue, yellow, or red path (more details in Sect.~\ref{secclasif}).}
\end{center}
\end{figure}

The workflow of HiLineThere is described in detail in Section \ref{HLT_section}. In Fig.~\ref{fig:WF_DETAIL}, we show a visual representation of the classifier, the part of this workflow used to determine the existence of diagnostic lines. 

\begin{figure*}
\sidecaption
\includegraphics[width=12cm]{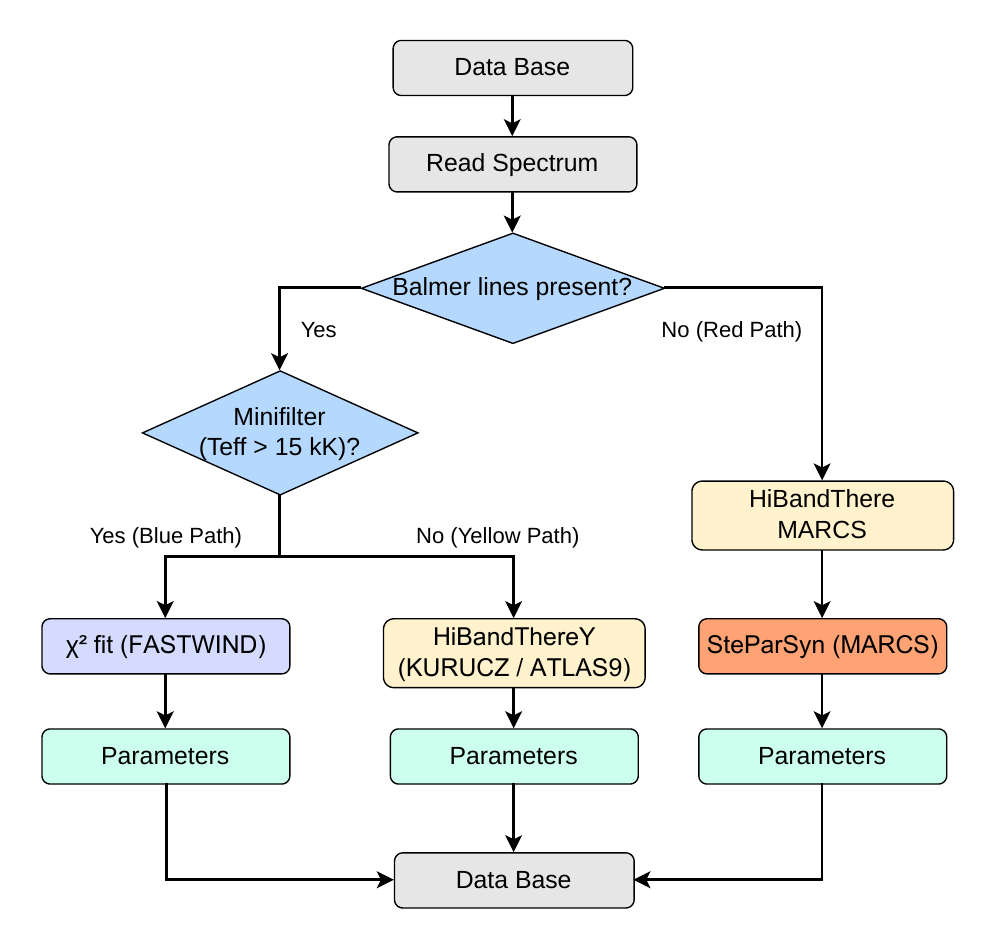}
\caption{\label{fig:WF_HLT} Science workflow, governed by a hierarchical classification. The pipeline first checks for the presence of Balmer lines to separate cool stars (red path) from hotter ones. For stars with Balmer lines, a 15\,000\,K minifilter further routes them into the blue path (analyzed via $\chi^2$ fitting and FASTWIND models) or the yellow path (analyzed via HiBandThereY and KURUCZ/ATLAS9 models). In the red path, HiBandThere computes the radial velocity before passing the spectrum to \textsc{SteParSyn}. At the end of all paths, the parameters are successfully stored in the database.}
\end{figure*}

The general workflow of the Astro+ procedures is illustrated in Fig.~\ref{fig:WF_HLT}, and is described in detail in Section~\ref{Science}.

\newpage
\section{Table of results}

\setlength{\LTcapwidth}{0.68\textwidth}
\begin{longtable}{@{}l|l|cc|cc|cc@{}}
\caption{\label{tab:listOB} Stellar parameters derived by Astro+ for the stars studied by \citet{holgado2018}.} \\ \hline\hline
\noalign{\smallskip}
&&\multicolumn{2}{c|}{Test 1} &\multicolumn{2}{c|}{Test 2}  &&\\
Star ID&Sptype  & $T_{\mathrm{eff}}$ & $\log \, \ensuremath{g}$ & $T_{\mathrm{eff}}$ & $\log \, \ensuremath{g}$  &  $V \sin\, i$ &  $V \zeta$   \\
&&(kK)&(dex)&(kK)&(dex)&(km\:s$^{-1}$)&(km\:s$^{-1}$)\\
\noalign{\smallskip}
\hline
\noalign{\smallskip}
\endfirsthead
\caption{continued.}\\
\hline\hline
\noalign{\smallskip}
&&\multicolumn{2}{c|}{Test 1} &\multicolumn{2}{c|}{Test 2}  &&\\
Star ID&Sptype  & $T_{\mathrm{eff}}$ & $\log \, \ensuremath{g}$ & $T_{\mathrm{eff}}$ & $\log \, \ensuremath{g}$  &  $V \sin\, i$ &  $V \zeta$   \\
&&(kK)&(dex)&(kK)&(dex)&(km\:s$^{-1}$)&(km\:s$^{-1}$)\\
\noalign{\smallskip}
\hline
\noalign{\smallskip}
\endhead
\noalign{\smallskip}
\hline
\noalign{\smallskip}
\hline
\endfoot
HD303311 & O6 V ((f))z & 38.9 $\pm$ 0.3 & 3.79 $\pm$ 0.09 & 38.9 $\pm$ 0.3 & 3.77 $\pm$ 0.08 & 48.5 $\pm$ 6.6 & 84.6 $\pm$ 10.5 \\
HD155889 & O9.5 IV & 36.0 $\pm$ 0.7 & 4.21 $\pm$ 0.06 & 35.6 $\pm$ 0.5 & 4.30 $\pm$ 0.10 & 29.6 $\pm$ 2.5 & 43.4 $\pm$ 2.7 \\
HD96715 & O4 V ((f))z & 44.0 $\pm$ 1.0 & 3.82 $\pm$ 0.07 & 43.9 $\pm$ 0.9 & 3.78 $\pm$ 0.07 & 64.0 $\pm$ 7.5 & 91.2 $\pm$ 15.3 \\
HD151804 & O8 Ia f & 31.0 $\pm$ 1.0 & 3.30 $\pm$ 0.20 & 31.0 $\pm$ 1.0 & 3.30 $\pm$ 0.20 & 67.5 $\pm$ 3.1 & 83.1 $\pm$ 3.6 \\
HD104565 & OC9.7 Iab & 29.0 $\pm$ 1.0 & 3.20 $\pm$ 0.10 & 29.4 $\pm$ 0.5 & 3.30 $\pm$ 0.10 & 79.0 $\pm$ 5.3 & 82.1 $\pm$ 6.6 \\
HD152405 & O9.7 II & 29.8 $\pm$ 0.8 & 3.20 $\pm$ 0.10 & 29.6 $\pm$ 0.5 & 3.19 $\pm$ 0.08 & 57.3 $\pm$ 3.8 & 68.1 $\pm$ 4.8 \\
HD93843 & O5 III (fc) & 38.0 $\pm$ 1.0 & 3.60 $\pm$ 0.20 & 38.2 $\pm$ 0.8 & 3.50 $\pm$ 0.10 & 64.5 $\pm$ 5.9 & 111.0 $\pm$ 5.7 \\
HD71304 & O9 II & 31.3 $\pm$ 0.4 & 3.19 $\pm$ 0.08 & 30.7 $\pm$ 0.4 & 3.09 $\pm$ 0.06 & 61.5 $\pm$ 4.6 & 92.4 $\pm$ 4.6 \\
HD322417 & O6.5 IV ((f)) & 37.2 $\pm$ 0.9 & 3.60 $\pm$ 0.10 & 37.0 $\pm$ 0.7 & 3.51 $\pm$ 0.08 & 72.8 $\pm$ 5.3 & 101.6 $\pm$ 5.7 \\
HD319699 & O5 V ((fc)) & 40.0 $\pm$ 0.7 & 3.79 $\pm$ 0.08 & 40.9 $\pm$ 0.8 & 3.81 $\pm$ 0.08 & 101.9 $\pm$ 3.9 & 62.5 $\pm$ 14.1 \\
HD94963 & O7 II (f) & 35.6 $\pm$ 0.9 & 3.50 $\pm$ 0.10 & 34.9 $\pm$ 0.6 & 3.38 $\pm$ 0.08 & 71.3 $\pm$ 4.9 & 89.2 $\pm$ 5.7 \\
CPD-592600 & O6 V ((f)) & 39.2 $\pm$ 0.9 & 3.80 $\pm$ 0.10 & 39.2 $\pm$ 0.9 & 3.80 $\pm$ 0.10 & 108.3 $\pm$ 11.7 & 138.4 $\pm$ 16.1 \\
HD154811 & OC9.7 Ib & 29.8 $\pm$ 0.6 & 3.21 $\pm$ 0.08 & 29.4 $\pm$ 0.5 & 3.19 $\pm$ 0.06 & 96.6 $\pm$ 5.8 & 91.7 $\pm$ 9.4 \\
HD168112 & O5 III (f) & 38.4 $\pm$ 0.9 & 3.59 $\pm$ 0.08 & 38.9 $\pm$ 0.8 & 3.60 $\pm$ 0.10 & 105.9 $\pm$ 6.5 & 94.1 $\pm$ 11.5 \\
HD93160 & O7 III ((f)) & 36.7 $\pm$ 0.7 & 3.62 $\pm$ 0.07 & 36.5 $\pm$ 0.5 & 3.62 $\pm$ 0.07 & 140.2 $\pm$ 4.2 & 92.3 $\pm$ 14.4 \\
HD152590 & O7.5 V z & 37.5 $\pm$ 0.5 & 3.85 $\pm$ 0.05 & 37.0 $\pm$ 0.7 & 3.80 $\pm$ 0.10 & 44.1 $\pm$ 1.8 & 67.1 $\pm$ 1.6 \\
HD93249 & O9 III & 33.1 $\pm$ 0.8 & 3.60 $\pm$ 0.20 & 33.0 $\pm$ 0.7 & 3.60 $\pm$ 0.20 & 56.6 $\pm$ 3.3 & 65.4 $\pm$ 3.4 \\
HD46202 & O9.2 V & 34.2 $\pm$ 0.4 & 3.90 $\pm$ 0.10 & 35.0 $\pm$ 1.0 & 4.10 $\pm$ 0.30 & 16.1 $\pm$ 1.5 & 35.8 $\pm$ 1.4 \\
CPD-592551 & O9 V & 34.7 $\pm$ 0.7 & 3.90 $\pm$ 0.10 & 35.9 $\pm$ 0.3 & 4.38 $\pm$ 0.06 & 127.8 $\pm$ 3.4 & 34.2 $\pm$ 19.4 \\
HD154643 & O9.7 III & 30.7 $\pm$ 0.7 & 3.46 $\pm$ 0.07 & 30.4 $\pm$ 0.5 & 3.46 $\pm$ 0.05 & 105.0 $\pm$ 2.3 & 67.4 $\pm$ 8.5 \\
HD156738 & O6.5 III (f) & 37.0 $\pm$ 1.0 & 3.70 $\pm$ 0.20 & 36.8 $\pm$ 0.9 & 3.60 $\pm$ 0.10 & 54.3 $\pm$ 8.3 & 104.1 $\pm$ 10.0 \\
HD91824 & O7 V ((f))z & 38.3 $\pm$ 0.7 & 3.90 $\pm$ 0.10 & 38.5 $\pm$ 0.5 & 3.91 $\pm$ 0.09 & 50.2 $\pm$ 2.6 & 65.7 $\pm$ 3.1 \\
HD12993 & O6.5 V ((f))Nstr & 38.0 $\pm$ 0.5 & 3.75 $\pm$ 0.09 & 38.0 $\pm$ 0.5 & 3.74 $\pm$ 0.07 & 79.7 $\pm$ 5.5 & 67.8 $\pm$ 9.4 \\
HD93128 & O3.5 V ((fc))z & 46.0 $\pm$ 1.0 & 3.84 $\pm$ 0.05 & 46.0 $\pm$ 1.0 & 3.84 $\pm$ 0.07 & 74.2 $\pm$ 10.2 & 84.0 $\pm$ 17.5 \\
HD57061 & O9 II & 32.2 $\pm$ 0.6 & 3.30 $\pm$ 0.10 & 33.0 $\pm$ 1.0 & 3.40 $\pm$ 0.20 & 54.3 $\pm$ 4.8 & 92.0 $\pm$ 4.7 \\
HD152424 & OC9.2 Ia & 29.9 $\pm$ 0.8 & 3.10 $\pm$ 0.10 & 29.8 $\pm$ 0.8 & 3.10 $\pm$ 0.10 & 55.8 $\pm$ 4.8 & 101.7 $\pm$ 4.7 \\
HD225146 & O9.7 Iab & 29.2 $\pm$ 0.7 & 3.20 $\pm$ 0.10 & 28.8 $\pm$ 0.7 & 3.23 $\pm$ 0.07 & 71.2 $\pm$ 7.1 & 79.7 $\pm$ 12.7 \\
HD17603 & O7.5 Ib (f) & 33.5 $\pm$ 0.7 & 3.30 $\pm$ 0.20 & 32.8 $\pm$ 0.4 & 3.21 $\pm$ 0.08 & 106.5 $\pm$ 3.6 & 107.4 $\pm$ 6.1 \\
HD10125 & O9.7 II & 29.6 $\pm$ 0.5 & 3.24 $\pm$ 0.05 & 29.6 $\pm$ 0.5 & 3.25 $\pm$ 0.05 & 105.9 $\pm$ 10.6 & 148.5 $\pm$ 14.0 \\
HD35619 & O7.5 V ((f)) & 37.0 $\pm$ 0.7 & 3.80 $\pm$ 0.10 & 36.9 $\pm$ 0.6 & 3.81 $\pm$ 0.09 & 42.2 $\pm$ 3.0 & 58.6 $\pm$ 3.1 \\
HD93146 & O7 V ((f)) & 38.4 $\pm$ 0.7 & 3.84 $\pm$ 0.09 & 38.2 $\pm$ 0.4 & 3.80 $\pm$ 0.10 & 58.7 $\pm$ 3.7 & 76.4 $\pm$ 3.9 \\
HD42088 & O6 V ((f))z & 39.2 $\pm$ 0.4 & 3.89 $\pm$ 0.08 & 38.6 $\pm$ 0.5 & 3.79 $\pm$ 0.08 & 57.9 $\pm$ 3.3 & 66.6 $\pm$ 4.0 \\
HD152723 & O6.5 III (f) & 37.2 $\pm$ 0.8 & 3.70 $\pm$ 0.10 & 37.2 $\pm$ 0.8 & 3.70 $\pm$ 0.10 & 68.1 $\pm$ 5.1 & 107.5 $\pm$ 5.4 \\
HD38666 & O9.5 V & 33.0 $\pm$ 0.8 & 3.70 $\pm$ 0.10 & 33.0 $\pm$ 1.0 & 3.80 $\pm$ 0.20 & 109.8 $\pm$ 7.2 & 98.7 $\pm$ 13.1 \\
HD47432 & O9.7 Ib & 29.1 $\pm$ 0.8 & 3.10 $\pm$ 0.10 & 29.2 $\pm$ 0.7 & 3.10 $\pm$ 0.10 & 98.7 $\pm$ 2.5 & 69.3 $\pm$ 7.7 \\
BD-114586 & O8 Ib (f) & 32.0 $\pm$ 1.0 & 3.30 $\pm$ 0.20 & 32.0 $\pm$ 0.7 & 3.12 $\pm$ 0.08 & 58.4 $\pm$ 3.9 & 94.1 $\pm$ 4.1 \\
HD191978 & O8 V & 34.7 $\pm$ 0.5 & 3.59 $\pm$ 0.08 & 34.7 $\pm$ 0.4 & 3.61 $\pm$ 0.08 & 59.0 $\pm$ 2.5 & 80.1 $\pm$ 2.8 \\
HD93222 & O7 V ((f)) & 36.0 $\pm$ 1.0 & 3.60 $\pm$ 0.10 & 35.6 $\pm$ 0.7 & 3.50 $\pm$ 0.10 & 53.1 $\pm$ 3.4 & 87.1 $\pm$ 3.5 \\
HD76968 & O9.2 Ib & 31.0 $\pm$ 0.7 & 3.30 $\pm$ 0.10 & 30.6 $\pm$ 0.7 & 3.20 $\pm$ 0.10 & 42.6 $\pm$ 3.8 & 83.3 $\pm$ 3.4 \\
HD298429 & O8.5 V & 33.7 $\pm$ 0.5 & 3.58 $\pm$ 0.09 & 33.6 $\pm$ 0.5 & 3.60 $\pm$ 0.10 & 88.6 $\pm$ 4.4 & 51.7 $\pm$ 18.7 \\
HD93028 & O9 IV & 36.0 $\pm$ 1.0 & 3.90 $\pm$ 0.20 & 35.5 $\pm$ 0.5 & 3.90 $\pm$ 0.10 & 30.1 $\pm$ 2.7 & 49.6 $\pm$ 2.5 \\
HD97848 & O8 V & 34.6 $\pm$ 0.5 & 3.49 $\pm$ 0.06 & 34.5 $\pm$ 0.5 & 3.51 $\pm$ 0.07 & 48.5 $\pm$ 2.6 & 63.1 $\pm$ 2.8 \\
HD151515 & O7 II (f) & 36.0 $\pm$ 1.0 & 3.60 $\pm$ 0.20 & 35.0 $\pm$ 1.0 & 3.40 $\pm$ 0.20 & 65.6 $\pm$ 7.5 & 100.9 $\pm$ 11.2 \\
HD123008 & ON9.2 Iab & 30.4 $\pm$ 0.7 & 3.20 $\pm$ 0.20 & 31.0 $\pm$ 1.0 & 3.30 $\pm$ 0.20 & 66.9 $\pm$ 4.5 & 93.0 $\pm$ 4.8 \\
HD57236 & O8.5 V & 39.0 $\pm$ 1.0 & 4.40 $\pm$ 0.10 & 38.1 $\pm$ 0.6 & 4.43 $\pm$ 0.04 & 29.8 $\pm$ 3.0 & 46.3 $\pm$ 3.4 \\
CPD-472963 & O5 I fc & 38.0 $\pm$ 1.0 & 3.50 $\pm$ 0.10 & 38.0 $\pm$ 1.0 & 3.50 $\pm$ 0.10 & 85.8 $\pm$ 5.5 & 93.0 $\pm$ 7.5 \\
HD46223 & O4 V ((f)) & 43.5 $\pm$ 0.9 & 3.80 $\pm$ 0.07 & 43.0 $\pm$ 0.7 & 3.76 $\pm$ 0.07 & 61.4 $\pm$ 7.0 & 90.5 $\pm$ 8.1 \\
HD303492 & O8.5 Ia f & 26.0 $\pm$ 5.0 & 2.90 $\pm$ 0.40 & 29.7 $\pm$ 0.8 & 2.90 $\pm$ 0.10 & 55.8 $\pm$ 4.9 & 87.0 $\pm$ 5.4 \\
HD68450 & O9.7 II & 30.3 $\pm$ 0.7 & 3.30 $\pm$ 0.07 & 29.9 $\pm$ 0.6 & 3.26 $\pm$ 0.07 & 50.6 $\pm$ 3.6 & 69.3 $\pm$ 4.1 \\
HD169582 & O6 Ia f & 37.0 $\pm$ 1.0 & 3.50 $\pm$ 0.20 & 37.0 $\pm$ 0.7 & 3.50 $\pm$ 0.07 & 98.1 $\pm$ 5.2 & 34.6 $\pm$ 24.4 \\
HD225160 & O8 Iab f & 33.0 $\pm$ 1.0 & 3.40 $\pm$ 0.10 & 32.7 $\pm$ 0.8 & 3.30 $\pm$ 0.20 & 77.0 $\pm$ 4.8 & 103.7 $\pm$ 5.5 \\
HD125241 & O8.5 Ib (f) & 30.5 $\pm$ 0.9 & 3.30 $\pm$ 0.10 & 32.0 $\pm$ 1.0 & 3.40 $\pm$ 0.20 & 92.5 $\pm$ 5.4 & 130.3 $\pm$ 6.6 \\
HD91572 & O6.5 V ((f))z & 38.1 $\pm$ 0.8 & 3.80 $\pm$ 0.10 & 38.6 $\pm$ 0.5 & 3.85 $\pm$ 0.07 & 43.7 $\pm$ 4.4 & 79.6 $\pm$ 4.7 \\
HD319702 & O8 III & 37.0 $\pm$ 1.0 & 3.90 $\pm$ 0.10 & 35.9 $\pm$ 0.8 & 3.74 $\pm$ 0.09 & 54.2 $\pm$ 5.2 & 83.3 $\pm$ 5.7 \\
HD93027 & O9.5 IV & 33.9 $\pm$ 0.3 & 3.89 $\pm$ 0.06 & 33.9 $\pm$ 0.3 & 3.89 $\pm$ 0.08 & 44.2 $\pm$ 3.3 & 63.3 $\pm$ 3.7 \\
HD303308 & O4.5 V ((fc)) & 43.0 $\pm$ 1.0 & 3.84 $\pm$ 0.05 & 43.1 $\pm$ 0.8 & 3.83 $\pm$ 0.06 & 85.9 $\pm$ 8.6 & 85.8 $\pm$ 15.8 \\
HD101223 & O8 V & 35.2 $\pm$ 0.4 & 3.59 $\pm$ 0.09 & 34.8 $\pm$ 0.4 & 3.53 $\pm$ 0.08 & 55.7 $\pm$ 4.5 & 64.1 $\pm$ 4.6 \\
HD93250 & O4 III (fc) & 44.0 $\pm$ 1.0 & 3.79 $\pm$ 0.08 & 43.9 $\pm$ 0.9 & 3.73 $\pm$ 0.06 & 69.5 $\pm$ 8.0 & 113.8 $\pm$ 11.4 \\
HD93204 & O5.5 V ((f)) & 39.0 $\pm$ 1.0 & 3.73 $\pm$ 0.08 & 40.0 $\pm$ 1.0 & 3.71 $\pm$ 0.08 & 112.0 $\pm$ 6.1 & 94.0 $\pm$ 11.6 \\
HD101190 & O6 IV ((f)) & 40.0 $\pm$ 0.7 & 3.87 $\pm$ 0.07 & 40.2 $\pm$ 0.7 & 3.89 $\pm$ 0.08 & 49.8 $\pm$ 4.9 & 75.8 $\pm$ 6.1 \\
HD96622 & O9.2 IV & 34.0 $\pm$ 0.7 & 3.80 $\pm$ 0.10 & 33.3 $\pm$ 0.5 & 3.70 $\pm$ 0.08 & 39.3 $\pm$ 3.4 & 61.9 $\pm$ 3.8 \\
HD154368 & O9.2 Iab & 30.0 $\pm$ 1.0 & 3.00 $\pm$ 0.10 & 29.7 $\pm$ 0.7 & 2.90 $\pm$ 0.10 & 62.4 $\pm$ 2.9 & 86.5 $\pm$ 2.9 \\
CPD-417733 & O9 IV & 34.2 $\pm$ 0.4 & 3.65 $\pm$ 0.08 & 34.0 $\pm$ 0.2 & 3.65 $\pm$ 0.05 & 54.0 $\pm$ 3.2 & 77.0 $\pm$ 3.5 \\
HD14947 & O4.5 I f & 37.6 $\pm$ 0.9 & 3.60 $\pm$ 0.10 & 39.5 $\pm$ 0.9 & 3.63 $\pm$ 0.07 & 96.5 $\pm$ 13.4 & 154.1 $\pm$ 23.3 \\
HD207538 & O9.7 IV & 31.9 $\pm$ 0.6 & 3.80 $\pm$ 0.10 & 31.2 $\pm$ 0.8 & 3.70 $\pm$ 0.20 & 31.6 $\pm$ 2.6 & 46.9 $\pm$ 3.1 \\
HD167633 & O6.5 V ((f)) & 37.9 $\pm$ 0.6 & 3.61 $\pm$ 0.08 & 38.2 $\pm$ 0.4 & 3.56 $\pm$ 0.07 & 134.1 $\pm$ 3.0 & 93.2 $\pm$ 9.0 \\
HD168076 & O4 III (f) & 44.0 $\pm$ 1.0 & 3.75 $\pm$ 0.08 & 44.0 $\pm$ 1.0 & 3.75 $\pm$ 0.08 & 64.3 $\pm$ 11.7 & 122.1 $\pm$ 22.9 \\
HD189957 & O9.7 III & 31.2 $\pm$ 0.9 & 3.50 $\pm$ 0.20 & 29.9 $\pm$ 0.3 & 3.33 $\pm$ 0.06 & 89.0 $\pm$ 1.8 & 62.3 $\pm$ 5.7 \\
HD94024 & O8 IV & 35.5 $\pm$ 0.7 & 3.61 $\pm$ 0.09 & 35.7 $\pm$ 0.5 & 3.56 $\pm$ 0.08 & 162.1 $\pm$ 2.7 & 52.8 $\pm$ 14.7 \\
HD114737 & O8.5 III & 35.3 $\pm$ 0.7 & 3.80 $\pm$ 0.10 & 34.9 $\pm$ 0.7 & 3.70 $\pm$ 0.10 & 55.5 $\pm$ 4.6 & 92.5 $\pm$ 4.7 \\
HD93129A & O2 I f* & 43.0 $\pm$ 2.0 & 3.68 $\pm$ 0.07 & 43.0 $\pm$ 3.0 & 3.70 $\pm$ 0.10 & 118.1 $\pm$ 5.5 & 145.6 $\pm$ 6.9 \\
HD218915 & O9.2 Iab & 30.6 $\pm$ 0.7 & 3.20 $\pm$ 0.10 & 30.1 $\pm$ 0.6 & 3.10 $\pm$ 0.09 & 63.5 $\pm$ 3.5 & 86.2 $\pm$ 3.7 \\
HD36486 & O9.5 II Nwk & 30.0 $\pm$ 0.7 & 3.30 $\pm$ 0.10 & 30.0 $\pm$ 0.5 & 3.32 $\pm$ 0.08 & 88.3 $\pm$ 7.0 & 147.2 $\pm$ 6.2 \\
HD152147 & O9.7 Ib Nwk & 29.5 $\pm$ 0.5 & 3.22 $\pm$ 0.07 & 29.3 $\pm$ 0.7 & 3.20 $\pm$ 0.09 & 59.8 $\pm$ 6.0 & 118.2 $\pm$ 6.2 \\
HD64568 & O3 V ((f*))z & 46.7 $\pm$ 0.8 & 3.91 $\pm$ 0.03 & 46.5 $\pm$ 0.9 & 3.89 $\pm$ 0.03 & 57.3 $\pm$ 8.9 & 77.6 $\pm$ 17.3 \\
HD163758 & O6.5 Ia fp & 35.1 $\pm$ 0.8 & 3.51 $\pm$ 0.09 & 35.5 $\pm$ 0.5 & 3.51 $\pm$ 0.09 & 79.5 $\pm$ 6.8 & 87.6 $\pm$ 8.8 \\
HD152249 & OC9 Iab & 31.0 $\pm$ 0.7 & 3.20 $\pm$ 0.10 & 31.0 $\pm$ 0.7 & 3.20 $\pm$ 0.10 & 64.8 $\pm$ 2.8 & 91.0 $\pm$ 2.9 \\
HD96264 & O9.5 III & 35.0 $\pm$ 1.0 & 3.90 $\pm$ 0.30 & 34.0 $\pm$ 1.0 & 3.80 $\pm$ 0.30 & 40.4 $\pm$ 3.6 & 54.5 $\pm$ 4.2 \\
HD156154 & O7.5 Ib (f) & 33.4 $\pm$ 0.9 & 3.20 $\pm$ 0.10 & 33.3 $\pm$ 0.5 & 3.18 $\pm$ 0.06 & 59.5 $\pm$ 4.2 & 95.3 $\pm$ 4.3 \\
HD46149 & O8.5 V & 36.9 $\pm$ 0.6 & 4.07 $\pm$ 0.08 & 36.7 $\pm$ 0.4 & 4.10 $\pm$ 0.10 & 34.4 $\pm$ 2.5 & 45.3 $\pm$ 2.9 \\
HD96946 & O6.5 III (f) & 38.0 $\pm$ 1.0 & 3.70 $\pm$ 0.10 & 37.4 $\pm$ 0.8 & 3.60 $\pm$ 0.10 & 68.6 $\pm$ 6.9 & 85.1 $\pm$ 8.8 \\
HD105056 & ON9.7 Ia e & 19.0 $\pm$ 1.0 & 2.20 $\pm$ 0.10 & 28.0 $\pm$ 1.0 & 3.00 $\pm$ 0.20 & 53.9 $\pm$ 3.7 & 61.3 $\pm$ 4.1 \\
HD135591 & O8 IV ((f)) & 35.0 $\pm$ 0.5 & 3.52 $\pm$ 0.08 & 34.4 $\pm$ 0.5 & 3.44 $\pm$ 0.08 & 58.1 $\pm$ 1.5 & 66.4 $\pm$ 2.2 \\
HD69464 & O7 Ib (f) & 35.3 $\pm$ 0.8 & 3.40 $\pm$ 0.10 & 35.2 $\pm$ 0.8 & 3.40 $\pm$ 0.10 & 64.3 $\pm$ 5.8 & 98.5 $\pm$ 7.3 \\
HD149038 & O9.7 Iab & 29.0 $\pm$ 1.0 & 3.20 $\pm$ 0.10 & 29.0 $\pm$ 1.0 & 3.10 $\pm$ 0.20 & 55.7 $\pm$ 2.8 & 77.6 $\pm$ 3.2 \\
HD14633 & ON8.5 V & 33.5 $\pm$ 0.5 & 3.53 $\pm$ 0.05 & 33.0 $\pm$ 1.0 & 3.60 $\pm$ 0.30 & 117.6 $\pm$ 4.1 & 97.6 $\pm$ 10.4 \\
HD75211 & O8.5 II ((f)) & 33.0 $\pm$ 1.0 & 3.30 $\pm$ 0.10 & 33.5 $\pm$ 0.7 & 3.30 $\pm$ 0.10 & 144.0 $\pm$ 2.2 & 85.2 $\pm$ 9.3 \\
HD193443 & O9 III & 32.1 $\pm$ 0.6 & 3.44 $\pm$ 0.09 & 31.4 $\pm$ 0.5 & 3.31 $\pm$ 0.08 & 64.8 $\pm$ 7.1 & 141.4 $\pm$ 6.2 \\
HD190864 & O6.5 III (f) & 36.5 $\pm$ 0.5 & 3.50 $\pm$ 0.07 & 36.4 $\pm$ 0.5 & 3.43 $\pm$ 0.07 & 49.3 $\pm$ 4.9 & 103.5 $\pm$ 4.3 \\
HD24431 & O9 III & 33.8 $\pm$ 0.6 & 3.60 $\pm$ 0.10 & 33.1 $\pm$ 0.3 & 3.48 $\pm$ 0.07 & 59.0 $\pm$ 3.9 & 79.3 $\pm$ 4.5 \\
HD46966 & O8.5 IV & 35.4 $\pm$ 0.5 & 3.70 $\pm$ 0.10 & 34.9 $\pm$ 0.3 & 3.64 $\pm$ 0.07 & 34.5 $\pm$ 3.7 & 72.6 $\pm$ 3.4 \\
HD242926 & O7 V z & 37.6 $\pm$ 0.5 & 3.91 $\pm$ 0.09 & 37.6 $\pm$ 0.5 & 3.92 $\pm$ 0.09 & 100.1 $\pm$ 2.1 & 35.4 $\pm$ 11.3 \\
HD192001 & O9.5 IV & 33.1 $\pm$ 0.9 & 3.70 $\pm$ 0.20 & 33.1 $\pm$ 0.9 & 3.80 $\pm$ 0.20 & 39.4 $\pm$ 4.1 & 63.7 $\pm$ 3.9 \\
HD15629 & O4.5 V ((fc)) & 40.0 $\pm$ 1.0 & 3.66 $\pm$ 0.08 & 40.0 $\pm$ 1.0 & 3.63 $\pm$ 0.09 & 63.4 $\pm$ 5.5 & 99.9 $\pm$ 5.4 \\
HD171589 & O7.5 II (f) & 35.5 $\pm$ 0.5 & 3.55 $\pm$ 0.09 & 35.6 $\pm$ 0.5 & 3.50 $\pm$ 0.10 & 95.9 $\pm$ 2.6 & 89.3 $\pm$ 4.9 \\
HD195592 & O9.7 Ia & 24.0 $\pm$ 4.0 & 2.60 $\pm$ 0.30 & 27.0 $\pm$ 2.0 & 2.80 $\pm$ 0.20 & 50.6 $\pm$ 5.1 & 97.7 $\pm$ 5.2 \\
HD191781 & ON9.7 Iab & 29.0 $\pm$ 2.0 & 3.40 $\pm$ 0.40 & 28.0 $\pm$ 2.0 & 3.30 $\pm$ 0.40 & 88.3 $\pm$ 10.7 & 116.8 $\pm$ 19.4 \\
HD188209 & O9.5 Iab & 30.2 $\pm$ 0.6 & 3.16 $\pm$ 0.09 & 30.2 $\pm$ 0.4 & 3.12 $\pm$ 0.08 & 57.9 $\pm$ 3.0 & 84.3 $\pm$ 3.3 \\
HD188001 & O7.5 Iab f & 32.9 $\pm$ 0.8 & 3.40 $\pm$ 0.10 & 33.1 $\pm$ 0.8 & 3.40 $\pm$ 0.10 & 61.5 $\pm$ 3.7 & 103.2 $\pm$ 3.8 \\
HD192639 & O7.5 Iab f & 34.0 $\pm$ 1.0 & 3.40 $\pm$ 0.10 & 34.0 $\pm$ 1.0 & 3.40 $\pm$ 0.20 & 84.0 $\pm$ 5.1 & 112.5 $\pm$ 5.2 \\
HD190429 & O4 I f & 38.0 $\pm$ 1.0 & 3.57 $\pm$ 0.08 & 38.0 $\pm$ 1.0 & 3.56 $\pm$ 0.08 & 83.3 $\pm$ 8.4 & 122.2 $\pm$ 9.6 \\
\end{longtable}
\tablefoot{Test~1 parameters were derived by using the same lines and weights employed by \citet{holgado2018}, while Test~2 parameters were obtained by letting the algorithm run without imposing any condition.}

\setlength{\LTcapwidth}{0.68\textwidth}
\begin{longtable}{l l | cc | cc}
\caption{Stellar parameters derived for the sample from \citet{Nieva2014}.\label{tab:nieva_results}}\\
\hline\hline
Star ID & Sptype & $T_{\mathrm{eff}}$ (kK) & $\log g$ (dex) & $V \sin i$ (km s$^{-1}$) & $V \zeta$ (km s$^{-1}$) \\
\hline
\endfirsthead
\caption{continued.}\\
\hline\hline
Star ID & Sptype & $T_{\mathrm{eff}}$ (kK) & $\log g$ (dex) & $V \sin i$ (km s$^{-1}$) & $V \zeta$ (km s$^{-1}$) \\
\hline
\endhead
\hline
\endfoot
HD35299 & B1.5 V & 27.8 $\pm$ 0.4 & 4.00 $\pm$ 0.20 & 8.1 $\pm$ 0.8 & 13.4 $\pm$ 1.1 \\
HD149438 & B0.2 V & 30.9 $\pm$ 0.9 & 4.00 $\pm$ 0.10 & 8.3 $\pm$ 0.5 & 10.5 $\pm$ 0.7 \\
HD37744 & B1.5 V & 26.0 $\pm$ 1.0 & 4.18 $\pm$ 0.04 & 38.2 $\pm$ 0.4 & 7.7 $\pm$ 2.6 \\
HD209008 & B3 III & 15.0 $\pm$ 0.25 & 3.50 $\pm$ 0.25 & 20.4 $\pm$ 1.9 & 13.2 $\pm$ 4.7 \\
HD36512 & B0 V & 32.6 $\pm$ 0.5 & 4.00 $\pm$ 0.10 & 12.9 $\pm$ 1.6 & 28.6 $\pm$ 1.6 \\
HD36285 & B2 V & 23.0 $\pm$ 2.0 & 4.10 $\pm$ 0.20 & 9.5 $\pm$ 1.1 & 15.9 $\pm$ 1.1 \\
HD29248 & B1.5 IV & 24.4 $\pm$ 0.7 & 3.80 $\pm$ 0.10 & 21.1 $\pm$ 0.9 & 33.6 $\pm$ 0.9 \\
HD205021 & B1 IV & 28.6 $\pm$ 0.7 & 4.00 $\pm$ 0.20 & 24.8 $\pm$ 0.9 & 29.5 $\pm$ 1.1 \\
HD36822 & B0.5 III & 31.0 $\pm$ 0.5 & 4.00 $\pm$ 0.10 & 21.7 $\pm$ 2.9 & 32.7 $\pm$ 4.5 \\
HD886 & B2 IV & 25.0 $\pm$ 1.0 & 4.10 $\pm$ 0.10 & 7.4 $\pm$ 0.7 & 14.1 $\pm$ 0.6 \\
HD61068 & B1 V & 28.0 $\pm$ 1.0 & 4.17 $\pm$ 0.09 & 12.0 $\pm$ 1.1 & 22.5 $\pm$ 1.0 \\
HD16582 & B2 IV & 24.0 $\pm$ 1.0 & 4.00 $\pm$ 0.10 & 8.1 $\pm$ 0.8 & 18.9 $\pm$ 0.6 \\
HD36960 & B0.7 V & 30.0 $\pm$ 1.0 & 4.10 $\pm$ 0.20 & 20.7 $\pm$ 1.9 & 36.9 $\pm$ 1.9 \\
HD37020 & B0.5 V & 28.9 $\pm$ 0.8 & 4.01 $\pm$ 0.08 & 54.5 $\pm$ 0.7 & 18.5 $\pm$ 4.6 \\
HD35039 & B2 IV & 22.0 $\pm$ 4.0 & 3.70 $\pm$ 0.40 & 9.4 $\pm$ 0.9 & 16.3 $\pm$ 0.8 \\
HD36959 & B1.5 V & 27.9 $\pm$ 0.6 & 4.21 $\pm$ 0.09 & 12.1 $\pm$ 0.5 & 12.6 $\pm$ 0.8 \\
HD37042 & B0.5 V & 29.0 $\pm$ 1.0 & 4.20 $\pm$ 0.10 & 33.3 $\pm$ 0.3 & 8.4 $\pm$ 2.1 \\
HD34816 & B0.5 V & 30.4 $\pm$ 0.5 & 4.10 $\pm$ 0.10 & 29.4 $\pm$ 3.5 & 24.5 $\pm$ 9.2 \\
\hline
\end{longtable}

\setlength{\LTcapwidth}{0.5\textwidth}
\begin{longtable}{@{}lcccc@{}}
\caption{Stellar parameters obtained for a sample of cool supergiants from \citet{Negueruela2018}, analyzed with the red path.\label{listRed}}\\
\hline\hline
\noalign{\smallskip}
Star ID & $T_{\mathrm{eff}}$ & e$T_{\mathrm{eff}}$  & $\log \, \ensuremath{g}$ & e$\log \, \ensuremath{g}$  \\
      & K  & K & dex& dex  \\
      \noalign{\smallskip}
\hline
\noalign{\smallskip}
\endfirsthead
\caption{continued.}\\
\hline\hline
\noalign{\smallskip}
Star ID & $T_{\mathrm{eff}}$ & e($T_{\mathrm{eff}})$  & $\log \, \ensuremath{g}$ & e($\log \, \ensuremath{g}$)  \\
\noalign{\smallskip}
\hline
\endhead
\noalign{\smallskip}
\hline
\endfoot
ASASSN-V J201151.18+342447.2    & 6200 & 300 & 1.3 & 0.5 \\
UCAC4 623-096276                & 4500 & 250 & 2.1 & 0.7 \\
2MASS J20115005+3425028         & 4670 & 220 & 2.0 & 0.7 \\
2MASS J20115643+3423580         & 6200 & 300 & 1.1 & 0.3 \\
2MASS J20115247+3424175         & 4700 & 300 & 1.8 & 0.8 \\
2MASS J20115645+3424222         & 6710 & 210 & 1.8 & 0.4 \\
IRAS 20100+3415                 & 3710 & 210 & 0.6 & 0.3 \\
IRAS 20099+3418                 & 3780 & 190 & 0.7 & 0.3 \\
2MASS J20121264+3420366         & 4790 & 220 & 1.5 & 0.8 \\
2MASS J20114858+3424420         & 4400 & 300 & 2.2 & 0.7 \\
UCAC4 622-094915                & 6100 & 400 & 1.6 & 0.6 \\
\hline

\end{longtable}
\twocolumn

\end{appendix}

\end{document}